\documentstyle[seceq,preprint,epsf]{jpsj}

\def\runtitle{Theory on Superconducting Transition from Pseudogap State} 
\def\runauthor{Youichi {\sc Yanase}, Takanobu {\sc Jujo}, Kosaku {\sc Yamada} }

\title{ % Don't Edit
Theory on Superconducting Transition from Pseudogap State 
} % Don't Edit

\author{Youichi {\sc Yanase}\footnote{E-mail: yanase@ton.scphys.kyoto-u.ac.jp},
Takanobu {\sc Jujo} and Kosaku {\sc Yamada}}

\inst{Department of Physics, Kyoto University, Kyoto 606-8502}

\recdate{March 22, 2000}

\abst
{
  The anomalous properties of High-$T_{{\rm c}}$ cuprates are investigated 
both in the normal state and in the superconducting state. 
 In particular, we pay attention to the pseudogap in the normal state and 
the phase transition from the pseudogap state to the superconducting state. 
  The pseudogap phenomena observed in cuprates are naturally understood 
as a precursor of the strong coupling superconductivity. 
 We have previously shown by using the self-consistent T-matrix 
calculation that the pseudogap is a result of the strong 
superconducting fluctuations which are accompanied by the strong 
coupling superconductivity in quasi-two dimensional systems 
[J. Phys. Soc. Jpn. {\bf68} (1999) 2999.].   
  We extend the scenario to the superconducting state. The close relation 
between the pseudogap state and the superconducting state is pointed out. 
  Once the superconducting phase transition occurs, 
the superconducting order parameter rapidly grows rather than the result of 
BCS theory. 
 With the rapid growth of the order parameter, 
the gap structure becomes sharp, while it is remarkably broad 
in the pseudogap state. 
 The characteristic energy scale of the gap does not change. 
 These results well explain the phase transition observed in the 
spectroscopic measurements. 
 Further, we calculate the magnetic and transport properties 
which show the pseudogap phenomena. 
 The comprehensive understanding of the NMR, the neutron scattering, 
the optical conductivity and the London penetration depth 
is obtained both in the 
pseudogap state and in the superconducting state. 
}

\kword
{High-$T_{{\rm c}}$ cuprates; Pseudogap; Superconducting state; 
Strong coupling superconductivity; \\
Superconducting fluctuation; Magnetic properties; Transport }

\begin{document}
\sloppy
\maketitle

\section{Introduction}

  The pseudogap phenomena and their related issues 
have been investigated for many years from various points of view. 
  They are considered to be key issues for the comprehensive 
understanding of the High-$T_{{\rm c}}$ superconductivity. 
 The pseudogap phenomena mean the suppression of the spectral weight near the 
Fermi energy without any long range order. 
 They are universal phenomena observed in various compounds of 
High-$T_{{\rm c}}$ cuprates in the under-doped region. 

  Many experiments such as the nuclear magnetic resonance (NMR),~\cite{rf:NMR} 
optical conductivity,~\cite{rf:homes} transport,~\cite{rf:transport} 
electronic specific heat,~\cite{rf:momono} 
angle-resolved photo-emission spectroscopy 
(ARPES),~\cite{rf:ARPES} tunneling spectroscopy,~\cite{rf:renner} 
and so on have indicated the existence of the pseudogap in the normal state 
High-$T_{{\rm c}}$ cuprates from optimally- to under-doped region. 

 Many scenarios for the pseudogap phenomena have been theoretically proposed.
 The resonating valence bond (RVB) theory attributes the pseudogap to the 
singlet pairing of spinons.~\cite{rf:tanamoto} 
 Some authors have proposed the magnetic scenarios based on the 
anti-ferromagnetic or SDW gap formation or its precursor.~\cite{rf:SDW} 
 The scenarios as a precursor of the superconductivity have also been 
proposed~\cite{rf:emery,rf:randeriareview} 
 The scenario based on the phase fluctuations has been proposed by 
Emery and Kivelson~\cite{rf:emery} and calculated by other 
authors.~\cite{rf:phase} 
 The importance of the strong coupling superconductivity has been 
proposed~\cite{rf:randeriareview,rf:haussman,rf:stintzing,rf:kobayashiNSR,rf:koikegamiNSR} 
on the basis of the well-known Nozi$\grave{{\rm e}}$res and Schmitt-Rink (NSR) 
theory.~\cite{rf:Nozieres,rf:tokumitu}. 
 Furthermore, the strong coupling superconductivity 
has been phenomenologically proposed 
by Geshkenbein {\it et al.}~\cite{rf:geshkenbein} 
with reference to the sign problem of the 
fluctuational Hall effect.~\cite{rf:aronov} 

 We have previously shown that the pseudogap phenomena are naturally 
understood by considering the resonance scattering due to 
the strong superconducting fluctuations.~\cite{rf:yanasePG}  
 The strong thermal fluctuation is an inevitable result 
of the strong coupling superconductivity 
in quasi-two dimensional systems. 
 The strength of the superconducting coupling is indicated by the ratio 
$T_{{\rm c}}^{{\rm MF}}/\varepsilon_{{\rm F}}$. Here, $\varepsilon_{{\rm F}}$ 
is the effective Fermi energy, and $T_{{\rm c}}^{{\rm MF}}$ is the 
superconducting critical temperature obtained by the mean field theory. 
 Since the effective Fermi energy $\varepsilon_{{\rm F}}$ is renormalized by 
the electron-electron correlation, the ratio 
$T_{{\rm c}}/\varepsilon_{{\rm F}}$ increases in the strongly correlated 
electron systems.  Therefore, the strong coupling superconductivity has a 
general importance for the superconductivity in the strongly correlated 
electron systems. 
 Moreover, it is natural to consider the strong coupling superconductivity 
in High-$T_{{\rm c}}$ cuprates because of their high critical 
temperature itself. 

 It should be noticed that our scenario is different from the NSR
theory.\cite{rf:randeriareview,rf:Nozieres,rf:haussman,rf:stintzing,rf:kobayashiNSR,rf:koikegamiNSR} 
 The NSR theory is based on the chemical potential shift by the creation 
of the pre-formed pairs. 
 In the NSR scenario, the tightly bound pre-formed pairs exist above 
$T_{{\rm c}}$, and condensate at $T_{{\rm c}}$. 
 In this case, the phase transition is the Bose condensation in the 
conventional sense. 
 However, the NSR scenario is justified in the low density limit, but not in 
the case of High-$T_{{\rm c}}$ cuprates. 
High-$T_{{\rm c}}$ cuprates should be regarded as rather high density systems 
because they are nearly half-filled on the lattice. 
 The effective Fermi energy $\varepsilon_{{\rm F}}$ 
is reduced by the electron correlation effects, not by the low density. 
 Therefore, the physical picture based on the NSR theory is inappropriate 
for the pseudogap phenomena in High-$T_{{\rm c}}$ cuprates. 

 Our scenario is based on the resonance 
scattering~\cite{rf:tchernyshyov,rf:janko} due to the thermal 
superconducting fluctuations. 
 The strong superconducting fluctuations 
have serious effects on the electronic state and give rise 
to the pseudogap phenomena in quasi-two dimensional 
systems.~\cite{rf:yanasePG,rf:jujo}  
 In our scenario, 
the chemical potential shift is small and the 
phase transition should be regarded as the same as usual superconductivity, 
not the Bose condensation in the sense of the NSR theory. 
 The electronic structure has been calculated by various authors on the 
basis of the scenario with the superconducting origin.~\cite{rf:randeriareview,rf:haussman,rf:yanasePG,rf:tchernyshyov,rf:janko,rf:micnas,rf:dagotto,rf:ichinomiya,rf:jujo,rf:kobayasi,rf:koikegami,rf:yanaseMG,rf:jujoyanase,rf:onoda}  
 The physical pictures in each calculation (the superconductivity or the 
Bose condensation) are classified by the electron density 
in each model. 
 We have explained the pseudogap phenomena on the basis of the
resonance scattering scenario, 
where we have used the self-consistent T-matrix calculation and the TDGL 
expansion.~\cite{rf:yanasePG} 
 We have effectually found the self-consistent solution which shows the 
pseudogap in the quasi-two dimensional high density systems and 
near the critical temperature. 
 The self-consistent calculation in high density systems  
is difficult near $T_{{\rm c}}$ because of the strong fluctuations. 
The important point for the calculation is explained in ref.20. 
 
 Our physical picture is properly consistent with the magnetic field 
dependences of the pseudogap phenomena measured by the high field 
NMR experiments.~\cite{rf:zheng,rf:gorny,rf:mitrovic,rf:eschrig,rf:zheng2} 
 The response of the pseudogap to the magnetic field is different between 
the under-doped~\cite{rf:zheng} and slightly over-doped 
cuprates~\cite{rf:zheng2}. 
 Our calculation gives a comprehensive explanation for the experimental 
results in all doping rate.~\cite{rf:yanaseMG}   
 We rather think that the comprehensive understanding in 
the phase diagram supports our scenario. 
 It is not clear whether the explanation based on the magnetic 
origin~\cite{rf:tanamoto,rf:SDW} may be consistent with the magnetic field 
dependences, especially near the optimally-doped region.

 The direct measurements of the electronic spectrum, such as 
ARPES~\cite{rf:ARPES} and tunneling spectroscopy\cite{rf:renner} have 
indicated the similarity between the pseudogap and the superconducting gap. 
 In particular, the same energy scale and the same momentum dependence 
between the two gaps have been indicated. 
 Thus, the electronic structure changes continuously 
from the pseudogap state to the superconducting state. 
 This important observations have strongly suggested 
that the pseudogap is a precursor of the superconductivity. 
 The same energy scale of the two gaps are self-evident 
in the NSR theory because the energy scale is the binding energy of the 
pre-formed bosons.  
 However, it is not so self-evident in our scenario and should be confirmed 
by the calculations. 
 Furthermore, the similarity and the difference between the pseudogap state 
and the superconducting state have been reported from the various experiments. 
 The comprehensive explanation for the characteristic properties of the 
phase transition is desired. 

 In this paper, we extend the self-consistent 
T-matrix calculation to the superconducting state. 
 Our main purpose in this paper is to understand the 
superconducting transition from the pseudogap state. 
 The close relation indicated by the spectroscopic 
experiments are confirmed by our calculation. 
 We furthermore investigate the magnetic and transport properties 
which show the pseudogap phenomena. 
 The pseudogap phenomena observed in each probe are explained 
by considering the characteristic momentum and frequency dependence of 
High-$T_{{\rm c}}$ cuprates. 
 We calculate their behaviors from the pseudogap state to 
the superconducting state. 
 Our results give qualitatively consistent understanding for 
the experimental results.

 This paper is constructed as follows. 
 In \S2, we give a model Hamiltonian adopted in this paper and explain the 
theoretical framework. In \S3, the calculated results about the single 
particle properties are shown. In \S4 and \S5, the magnetic properties and 
the transport properties are investigated, respectively. 
In \S6, we summarize the obtained results and give some discussions.

\section{Theoretical Framework}

 In this section, we describe the theoretical framework in this paper. 
 Hereafter, we adopt the unit $\hbar=c=k_{{\rm B}}=1$. 
 Our calculation is based on the self-consistent T-matrix calculation. 
We have used this approximation in the pervious paper in order to describe 
the normal state pseudogap.~\cite{rf:yanasePG} 
 First, we explain the model Hamiltonian and review the formalism 
describing the normal state pseudogap. 
  We adopt the following two-dimensional model Hamiltonian which has a 
$d_{x^2-y^2}$-wave superconducting ground state, 
\begin{eqnarray}
  \label{eq:model}
  H =  \sum_{\mbox{\boldmath$k$},s}  \varepsilon_{\mbox{{\scriptsize \boldmath$k$}}} 
c_{\mbox{{\scriptsize \boldmath$k$}},s}^{\dag} c_{\mbox{{\scriptsize \boldmath$k$}},s}  
 +  \sum_{\mbox{\boldmath$k$},\mbox{\boldmath$k'$},\mbox{\boldmath$q$}} 
V_{\mbox{{\scriptsize \boldmath$k$}}-\mbox{{\scriptsize \boldmath$q$}}/2,
\mbox{{\scriptsize \boldmath$k'$}}-\mbox{{\scriptsize \boldmath$q$}}/2} 
c_{\mbox{{\scriptsize \boldmath$q$}}-\mbox{{\scriptsize \boldmath$k'$}},\downarrow}^{\dag} 
c_{\mbox{{\scriptsize \boldmath$k'$}},\uparrow}^{\dag} 
c_{\mbox{{\scriptsize \boldmath$k$}},\uparrow} 
c_{\mbox{{\scriptsize \boldmath$q$}}-\mbox{{\scriptsize \boldmath$k$}},\downarrow}, 
\end{eqnarray}
  where 
$ V_{\mbox{{\scriptsize \boldmath$k$}},\mbox{{\scriptsize \boldmath$k'$}}} $ 
is the $d_{x^2-y^2}$-wave separable pairing interaction, 
\begin{eqnarray}
  \label{eq:d-wave}
  V_{\mbox{{\scriptsize \boldmath$k$}},\mbox{{\scriptsize \boldmath$k'$}}} = 
g \varphi_{\mbox{{\scriptsize \boldmath$k$}}} \varphi_{\mbox{{\scriptsize \boldmath$k'$}}}, \\
  \varphi_{\mbox{{\scriptsize \boldmath$k$}}} = \cos k_{x}-\cos k_{y}. 
\end{eqnarray}

 Here, $g$ is negative and $ \varphi_{ \mbox{{\scriptsize \boldmath$k$}} } $ 
is the $d_{x^2-y^2}$-wave form factor. 
  We consider the dispersion $\varepsilon_{\mbox{{\scriptsize \boldmath$k$}}}$ given by the 
tight-binding model for a square lattice including the nearest- and 
next-nearest-neighbor hopping $t$, $t'$, respectively, 
\begin{eqnarray}
 \label{eq:dispersion}
    \varepsilon_{\mbox{{\scriptsize \boldmath$k$}}} = 
                  -2 t (\cos k_{x} +\cos k_{y}) + 
                  4 t' \cos k_{x} \cos k_{y} - \mu. 
\end{eqnarray}

 We fix the lattice constant $ a = 1$. 
 We adopt $t=0.5 $ and $t'=0.45 t$. These parameters well reproduce 
the Fermi surface of the typical High-$T_{{\rm c}}$ cuprates,
${\rm Y}{\rm Ba}_{2}{\rm Cu}_{3}{\rm O}_{6+\delta}$ and 
${\rm Bi}_{2}{\rm Sr}_{2}{\rm Ca}{\rm Cu}_{2}{\rm O}_{8+\delta}$. 
 We choose the chemical potential $\mu$ so that the filling $n=0.9$, which  
corresponds to the hole doping $\delta=0.1$. 
 The above Hamiltonian is an effective model in which the paring interaction 
affects the renormalized quasi-particles. 
 Since the energy scale in the above model is the renormalized Fermi energy, 
the relatively large paring interaction is considered. 
 The realistic energy scale for cuprates 
is obtained by considering the renormalization 
as $\cong 1/10$ which is a relevant order in $d$-electron systems. 

 Actually, the origin of the pairing interaction should be considered to be 
the anti-ferromagnetic spin fluctuations.~\cite{rf:monthoux,rf:moriya} 
 There are studies dealing with the pairing interaction arising from 
the spin fluctuations on the basis of the fluctuation exchange (FLEX) 
approximation.~\cite{rf:dahm,rf:koikegami} 
 However, the details of the pairing interaction do not seriously affect the 
pseudogap phenomena as a precursor of the $d_{x^{2}-y^{2}}$-wave 
superconductivity. 
 There is a feedback effect on the pairing interaction 
arising from the pseudogap. 
 The pseudogap suppresses the low frequency component of the spin 
fluctuations. 
 However, the pairing interaction is mainly caused by 
the high frequency component compared with the energy scale of 
the superconducting fluctuations. 
 Therefore, we can neglect the feedback effect on the pairing interaction  
and start from the model with a fixed attractive interaction for simplicity. 
 There is no qualitative difference about the pseudogap resulting from the 
superconducting fluctuations and the phase transition from the pseudogap 
state.

 As an effect of the superconducting fluctuations, the corrections on 
the two-body correlation functions have been studied 
for a long time in the weak coupling limit. They are well known as  
the Aslamazov-Larkin term (AL term)~\cite{rf:AL} and the Maki-Thompson term 
(MT term).~\cite{rf:MT} 
 On the other hand, the superconducting fluctuations more seriously affect 
the one-particle electronic state in the strong coupling region. 
 Especially, the pseudogap phenomena are caused.   

 Generally, the superconducting fluctuations are expressed by the T-matrix 
which is represented by the ladder diagrams in the particle-particle 
channel (Fig. 1(a)). 
\begin{figure}[htbp]
 \begin{center} 
   \epsfysize=3cm
$$\epsffile{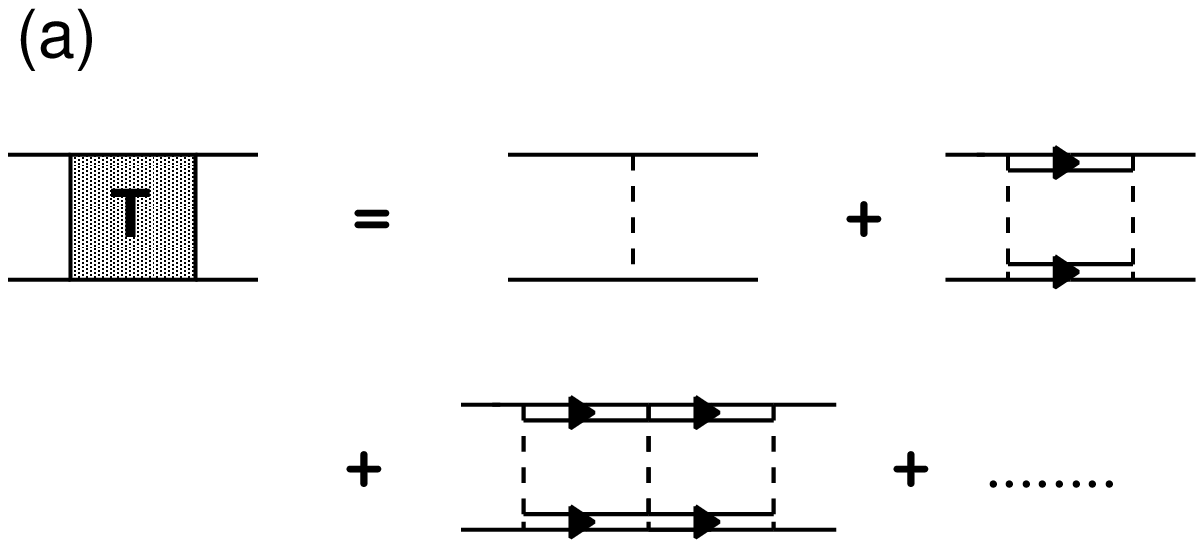}
%    \epsfile{file=vertexnormal.eps,height=3cm}
\hspace{5mm}
   \epsfysize=2cm
\epsffile{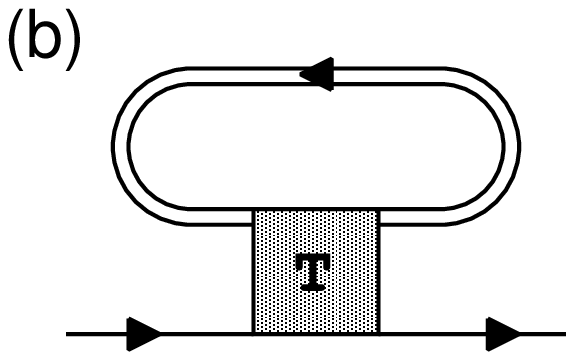}
%    \epsfile{file=diagramnormal.eps,height=2cm}
\hspace{5mm}
   \epsfysize=2cm
\epsffile{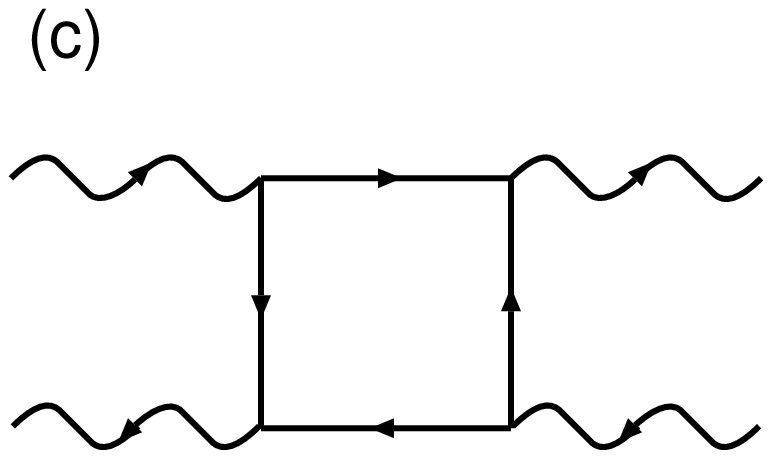}$$
%    \epsfile{file=critical.eps,height=2cm}
    \caption{(a) The T-matrix in the normal state.
             The dashed lines represent the attractive interaction. 
             The double solid lines represent the propagator of 
             the fermions. 
             (b) The self-energy in the normal state calculated by the 
              self-consistent T-matrix approximation. 
             (c) The diagram representing the mode coupling effect in the 
             lowest order. The wavy and solid lines represent 
             the propagator of the fluctuating Cooper pairs and that of the 
             fermions, respectively.
             }           
 \end{center}
\end{figure}
\begin{eqnarray}
  \label{eq:t-matrix}
  T(\mbox{\boldmath$q$},{\rm i} \Omega_{n}) & = & 
  [g^{-1} + \chi(\mbox{\boldmath$q$},{\rm i} \Omega_{n})]^{-1}, 
\\
  \chi(\mbox{\boldmath$q$},{\rm i} \Omega_{n}) & = & 
   T \sum_{\mbox{\boldmath$k'$},\omega_{m}} 
   {\mit{\it G}} (\mbox{\boldmath$k'$},{\rm i} \omega_{m}) 
   {\mit{\it G}} (\mbox{\boldmath$q$}-\mbox{\boldmath$k'$},
   {\rm i} \Omega_{n} - {\rm i} \omega_{m})
   \varphi_{\mbox{{\scriptsize \boldmath$k'$}} - 
            \mbox{{\scriptsize \boldmath$q$}}/2}^{2}.
\end{eqnarray}
 Here, $\omega_{m} = 2 \pi (m+\frac{1}{2}) T$ and $\Omega_{n}= 2 \pi n T$ are 
the fermionic and bosonic Matsubara frequencies, respectively. 
 The Green function ${\mit{\it G}}$ is expressed as  
${\mit{\it G}} (\mbox{\boldmath$k$}, {\rm i} \Omega_{n}) = 
(\omega - \varepsilon_{\mbox{{\scriptsize \boldmath$k$}}} - 
{\mit{\it \Sigma}} (\mbox{\boldmath$k$}, {\rm i} \Omega_{n}))^{-1}$. 
 The self-energy ${\mit{\it \Sigma}}$ is given by the self-consistent 
T-matrix calculation, (Fig. 1(b))  
\begin{eqnarray}
 \label{eq:selfenergymatubara}
  {\mit{\it \Sigma}} (\mbox{\boldmath$k$}, {\rm i} \omega_{m})  = 
  T \sum_{\mbox{\boldmath$q$},{\rm i} \Omega_{n}}
  T (\mbox{\boldmath$q$},{\rm i} \Omega_{m}) 
  {\mit{\it G}} (\mbox{\boldmath$q$}-\mbox{\boldmath$k$}, 
   {\rm i} \Omega_{n} - {\rm i} \omega_{m})
 \varphi_{\mbox{{\scriptsize \boldmath$k$}}-
          \mbox{{\scriptsize \boldmath$q$}}/2}^{2}.    
\end{eqnarray}
 The form factor $\varphi_{\mbox{{\scriptsize \boldmath$k$}}}$ in 
the scattering vertex gives rise to the $d_{x^{2}-y^{2}}$-wave shape of 
the pseudogap. 

 When  $ 1 + g \chi_{0}(\mbox{\boldmath$0$},0) = 0 $, 
the pair correlation function diverges and 
the superconducting transition occurs. 
 This is the Thouless criterion which is equivalent to  
the BCS theory in the weak coupling limit.~\cite{rf:Nozieres} 
 Analytically continued T-matrix $ T(\mbox{\boldmath$q$},\Omega) $ can be 
regarded as a propagator of the fluctuating Cooper pairs.

 It should be noticed that the self-consistently calculated T-matrix includes 
the renormalization effects of the superconducting fluctuations through the 
self-energy ${\mit{\it \Sigma}}^{\rm R} (\mbox{\boldmath$k$}, \omega)$. 
 The renormalization effects include the mode coupling 
effect.~\cite{rf:yanaseMG} 
 The forth order term in the Ginzburg-Landau action is expressed by the 
diagram shown in Fig. 1(c), which indicates the repulsive interaction 
between the fluctuating Cooper pairs (that is the lowest order mode coupling 
term). 
 The effect of the mode coupling term is included 
in the self-consistent T-matrix calculation at least 
in the Hartree-Fock approximation. 
 Thus, the self-consistent T-matrix calculation is a method 
introducing the criticality of the superconducting fluctuations. 
 Furthermore, the microscopic renormalization effects are included 
through the single particle properties.~\cite{rf:yanasePG} 
 As is explained below, the renormalization effects enhance the scattering 
vertex originated from the superconducting fluctuations and accelerate the 
pseudogap formation. 
 The critical temperature 
$T_{{\rm c}}$ is reduced by the fluctuations.~\cite{rf:yanasePG} 
 The reduced $T_{{\rm c}}$ is due to the reduced density of state (DOS) by 
the pseudogap. The reduced $T_{{\rm c}}$ can be regarded as a result of the 
wide critical region, simultaneously. 
 The reduction is remarkable in the strong coupling superconductivity, 
while it is neglected in the weak coupling one.

 In the previous paper, we have expanded 
$ T^{-1}(\mbox{\boldmath$q$},\Omega) $ around  
$ \mbox{\boldmath$q$} = \Omega = 0 $.~\cite{rf:yanasePG} 
 This expansion corresponds to the time-dependent-Ginzburg-Landau (TDGL) 
expansion,  
\begin{eqnarray}
  \label{eq:tdgl}
   T(\mbox{\boldmath$q$},\omega) = \frac{g}
   {t_{0} + b \mbox{\boldmath$q$}^{2} - (a_{1}+{\rm i}a_{2}) \Omega}. 
\end{eqnarray}

 The detailed properties of the TDGL parameters are discussed in ref. 20. 
 The outline is the following. The parameter 
$ t_{0} = 1 + g \chi_{0}(\mbox{\boldmath$0$},0) $ represents the distance to 
the phase transition, and is sufficiently small near $T_{{\rm c}}$. 
 The parameter $ b $ is generally related to the coherence length $\xi_{0}$, 
$b \propto \xi_{0}^{2}$. 
 The small $ b $ generally means the strong fluctuations. 
 The parameter $a_{2}$ expresses the time scale of the fluctuations. 
 Roughly speaking, the parameters $a_{2}$ and $b$ are described as, 
$a_{2} \propto \rho_{{\rm d}}(0)/T$ and $b \propto \rho_{{\rm d}}(0)/T^{2}$. 
 Here, $ \rho_{{\rm d}}(\varepsilon) $ is the effective density of states 
for the $ d_{x^{2}-y^{2}} $-wave symmetry.~\cite{rf:yanasePG}
 Because of the high critical temperature $T_{{\rm c}}$ and the 
renormalization effect by the pseudogap, both $a_{2}$ and $b$ are strongly 
reduced in the strong coupling superconductivity. 
 These features of the TDGL parameters indicate that the scattering vertex 
due to the superconducting fluctuations is strongly 
enhanced.
 The parameter $ a_{1} $ is determined by the particle-hole asymmetry 
of each system.~\cite{rf:ebisawa} 
 High-$T_{{\rm c}}$ cuprates have a large value of 
$ a_{1} $ because of their strongly asymmetric band structure. 
 Moreover, $ a_{1} $ is not so reduced even in case of the strong coupling 
superconductivity. 
 The extreme situation, $ |a_{1}| \ge  a_{2} $ can be realized near 
$T_{{\rm c}}$. In this case, the T-matrix has a propagative character. 
 However, it is a character of the collective mode and does not mean the 
tightly bound pairs supposed in the NSR scenario~\cite{rf:Nozieres}. 
 The finite $ a_{1} $ induces the asymmetry of the T-matrix. 
 Actually, the particle-hole asymmetry is not necessary but advantageous to 
the pseudogap state in the self-consistent solution.~\cite{rf:yanasePG}  
 The negative $ a_{1} $ is obtained in the optimally- and  under-doped region. 
 Thus, the superconducting fluctuations have a hole-like character 
even in the strong coupling superconductivity, 
contrary to the phenomenological assumption in ref. 18. 

 Thus, the superconducting fluctuations are strong in case of the strong 
coupling superconductivity. 
 Such strong fluctuations in quasi-two dimensional systems 
give rise to the anomalous properties of the 
self-energy compared with the conventional Fermi liquid theory. 
 The real part of the self-energy has a positive slope and the imaginary part 
has a large absolute value near the Fermi energy.~\cite{rf:yanasePG}  
 The large imaginary part suppresses the low frequency spectral weight, and 
the gap-like feature of the spectral weight is obtained. 
 This is the pseudogap, and is an effect of the resonance scattering 
due to the superconducting fluctuations.

 In this paper, we do not use the TDGL expansion with reference to the Landau 
singularity in the ordered state.~\cite{rf:tsuneto} 
 We explicitly calculate the T-matrix around $ \mbox{\boldmath$q$} = \Omega 
= 0 $. The obtained results confirm the above behaviors of the T-matrix 
including the renormalization effects. 
 The pseudogap state is obtained similarly. 
 We restrict the integrated area for $\mbox{\boldmath$q$}$ and $\Omega$, as is 
done in ref. 20. Namely, the meaningful region as a superconducting 
fluctuation is picked up. 
 There is no qualitative difference by the details of the calculation. 
 
 Here, we extend the self-consistent T-matrix calculation to the 
superconducting state. 
 The T-matrix is described as the following $ 2 \times 2 $ matrix in the 
superconducting state. (Fig. 2(a)) 
\begin{eqnarray}
  \mbox{\boldmath$T$}(\mbox{\boldmath$q$},{\rm i} \Omega_{n}) & = &
   [g^{-1} \mbox{\boldmath$1$} + 
     \mbox{\boldmath$\chi$}(\mbox{\boldmath$q$}, {\rm i} \Omega_{n})]^{-1}, \\
  \mbox{\boldmath$\chi$}(\mbox{\boldmath$q$},{\rm i} \Omega_{n}) & = &
  \left(
    \begin{array}{cc}
    K(\mbox{\boldmath$q$},{\rm i} \Omega_{n}) & 
    L(\mbox{\boldmath$q$},{\rm i} \Omega_{n}) \\
    L^{*}(\mbox{\boldmath$q$},{\rm i} \Omega_{n}) &
    K(\mbox{-\boldmath$q$},-{\rm i} \Omega_{n}) \\
    \end{array}
    \right),
\end{eqnarray}
  where, 
\begin{eqnarray}
K(\mbox{\boldmath$q$},{\rm i} \Omega_{n}) & =  & 
T \sum_{\mbox{\boldmath$k'$},\omega_{m}} {\mit{\it G}} (\mbox{\boldmath$k'$},
{\rm i} \omega_{m}) {\mit{\it G}} (\mbox{\boldmath$q$}-\mbox{\boldmath$k'$},
{\rm i} \Omega_{n} - {\rm i} \omega_{m})
\varphi_{\mbox{{\scriptsize \boldmath$k'$}}-\mbox{{\scriptsize \boldmath$q$}}/2}^{2}, 
\\
L(\mbox{\boldmath$q$},{\rm i} \Omega_{n}) & =  & 
- T \sum_{\mbox{\boldmath$k'$},\omega_{m}} {\mit{\it F}} (\mbox{\boldmath$k'$},
{\rm i} \omega_{m}) {\mit{\it F}} (\mbox{\boldmath$q$}-\mbox{\boldmath$k'$},
{\rm i} \Omega_{n} - {\rm i} \omega_{m})
\varphi_{\mbox{{\scriptsize \boldmath$k'$}}-\mbox{{\scriptsize \boldmath$q$}}/2}^{2},  
\end{eqnarray}
 and $\mbox{\boldmath$1$}$ is the unit matrix.

 Here, $ {\mit{\it G}} (\mbox{\boldmath$k$},{\rm i} \omega_{m}) $ and 
$ {\mit{\it F}} (\mbox{\boldmath$k$},{\rm i} \omega_{m}) $ are 
normal and anomalous Green functions including the self-energy 
$ {\mit{\it \Sigma}} (\mbox{\boldmath$k$}, {\rm i} \omega_{m}) $, 
respectively. 
\begin{eqnarray}
  {\mit{\it G}} (\mbox{\boldmath$k$},{\rm i} \omega_{m}) & = & 
 \frac{{\rm i} \omega_{m} + \varepsilon_{\mbox{{\scriptsize \boldmath$k$}}} + 
   {\mit{\it \Sigma}} (-\mbox{\boldmath$k$}, -{\rm i} \omega_{m})}
 {[{\rm i} \omega_{m} - \varepsilon_{\mbox{{\scriptsize \boldmath$k$}}} - 
   {\mit{\it \Sigma}} (\mbox{\boldmath$k$}, {\rm i} \omega_{m})]
   [{\rm i} \omega_{m} + \varepsilon_{\mbox{{\scriptsize \boldmath$k$}}} + 
   {\mit{\it \Sigma}} (-\mbox{\boldmath$k$}, -{\rm i} \omega_{m})]
  - \Delta_{\mbox{{\scriptsize \boldmath$k$}}}^{2}}, 
\\
   {\mit{\it F}} (\mbox{\boldmath$k$},{\rm i} \omega_{m}) & = & 
 \frac{- \Delta_{\mbox{{\scriptsize \boldmath$k$}}}}
 {[{\rm i} \omega_{m} - \varepsilon_{\mbox{{\scriptsize \boldmath$k$}}} - 
   {\mit{\it \Sigma}} (\mbox{\boldmath$k$}, {\rm i} \omega_{m})]
   [{\rm i} \omega_{m} + \varepsilon_{\mbox{{\scriptsize \boldmath$k$}}} + 
   {\mit{\it \Sigma}} (-\mbox{\boldmath$k$}, -{\rm i} \omega_{m})]
  - \Delta_{\mbox{{\scriptsize \boldmath$k$}}}^{2}}. 
\end{eqnarray}

 The normal self-energy 
$ {\mit{\it \Sigma}} (\mbox{\boldmath$k$}, {\rm i} \omega_{m}) $ is given by 
the self-consistent T-matrix approximation (Fig. 2(b)). 
\begin{eqnarray}
 \label{eq:selfenergymatubara}
  {\mit{\it \Sigma}} (\mbox{\boldmath$k$}, {\rm i} \omega_{m})  = 
  T \sum_{\mbox{\boldmath$q$},{\rm i} \Omega_{n}}
  T_{11} (\mbox{\boldmath$q$},{\rm i} \Omega_{m}) 
  {\mit{\it G}} (\mbox{\boldmath$q$}-\mbox{\boldmath$k$}, 
   {\rm i} \Omega_{n} - {\rm i} \omega_{m})
    \varphi_{\mbox{{\scriptsize \boldmath$k$}}-
             \mbox{{\scriptsize \boldmath$q$}}/2}^{2}.    
\end{eqnarray}
 Here, the trivial Hartree-Fock term is excluded (Fig. 2(c)). 

 The $d$-wave order parameter $\Delta_{\mbox{{\scriptsize \boldmath$k$}}}= 
\Delta \varphi_{\mbox{{\scriptsize \boldmath$k$}}}$ is determined 
by the gap equation. 
\begin{eqnarray}
 \label{eq:anormalousselfenergy}
  \Delta_{\mbox{{\scriptsize \boldmath$k$}}} = -g T 
  \sum_{\mbox{\boldmath$k'$},{\rm i} \Omega_{m}}
  {\mit{\it F}} (\mbox{\boldmath$k'$},{\rm i} \omega_{m})
  \varphi_{\mbox{{\scriptsize \boldmath$k'$}}} 
  \varphi_{\mbox{{\scriptsize \boldmath$k$}}}. 
\end{eqnarray}
 The effects of the fluctuations on the gap equation 
are included in the normal self-energy in 
${\mit{\it F}}(\mbox{\boldmath$k$},{\rm i} \omega_{m})$. 
\begin{figure}[htbp]
 \begin{center} 
   \epsfysize=4cm
$$\epsffile{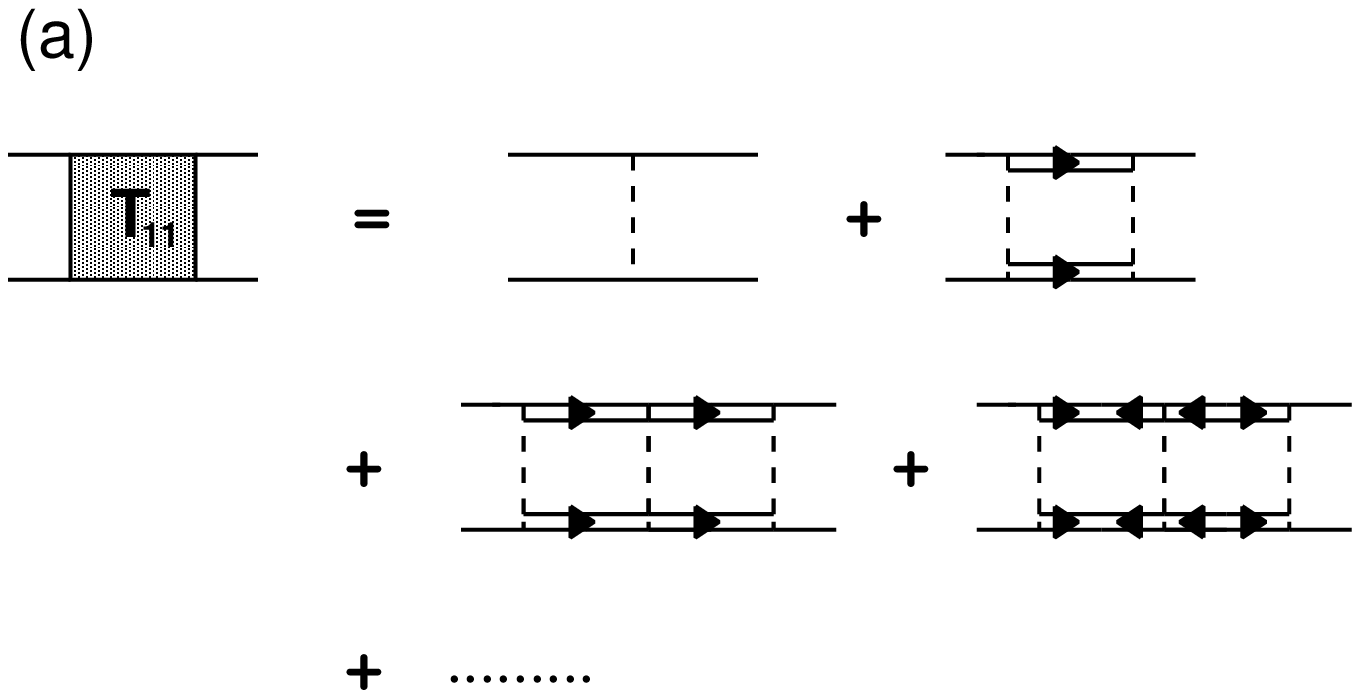}
%    \epsfile{file=vertex.eps,height=4cm}
\hspace{7mm}
   \epsfysize=2cm
\epsffile{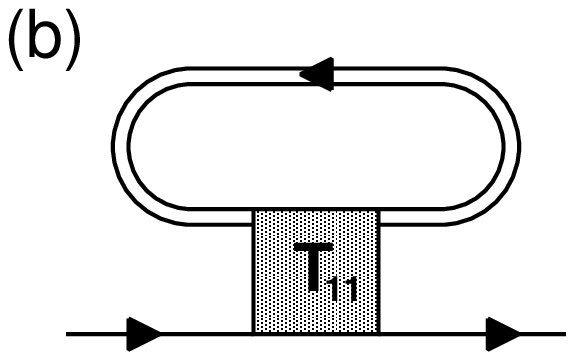}
%    \epsfile{file=diagram.eps,height=2cm}
\hspace{7mm}
   \epsfysize=2cm
\epsffile{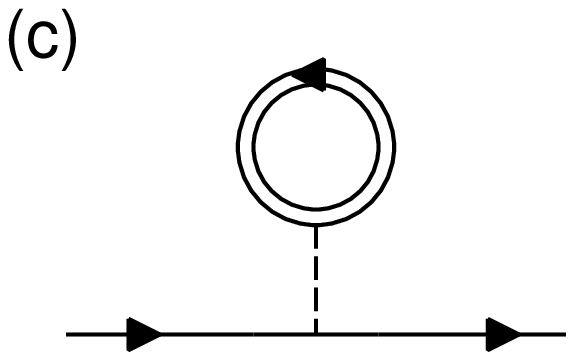}$$
%    \epsfile{file=Hartree.eps,height=2cm}
    \caption{(a) The diagonal component of the T-matrix in the superconducting 
             state.
             The double solid lines represent the normal and anomalous Green 
             functions of the fermions. 
             (b) The normal self-energy calculated by the self-consistent 
             T-matrix approximation. 
             (c) The Hartree-Fock term which we exclude. 
             }             
 \end{center}
\end{figure}

 The T-matrix corresponds to the propagator of the pair field 
$ \eta (\mbox{\boldmath$q$},{\rm i} \Omega_{n}) = \Delta 
(\mbox{\boldmath$q$},{\rm i} \Omega_{n}) - \Delta $.~\cite{rf:engelbrecht} 
 By describing the pair field by the amplitude mode 
$\lambda (\mbox{\boldmath$q$},{\rm i} \Omega_{n})$ and the phase mode   
$\theta (\mbox{\boldmath$q$},{\rm i} \Omega_{n})$ as 
$ \eta (\mbox{\boldmath$q$},{\rm i} \Omega_{n}) = 
\lambda (\mbox{\boldmath$q$},{\rm i} \Omega_{n}) + 
{\rm i}  \theta (\mbox{\boldmath$q$},{\rm i} \Omega_{n}) $, 
 the effective Gaussian action for the pair field is expressed as, 
\begin{eqnarray}
 S_{2} (\lambda,\theta) = \frac{1}{2} T \sum_{\mbox{\boldmath$q$},{\rm i} \Omega
_{n}}
(\lambda^{*},
 \theta^{*})
  \left(
    \begin{array}{cc}
    1/|g| - K_{+} 
    - L & 
    - {\rm i} K_{-} \\
      {\rm i} K_{-} &
    1/|g| - K_{+}
    + L \\
    \end{array}
    \right)
  \left(
    \begin{array}{c}
    \lambda \\
    \theta \\
    \end{array}
    \right).
\end{eqnarray}

 Here, 
$K_{+} = (K(\mbox{\boldmath$q$},{\rm i} \Omega_{n})+K(-\mbox{\boldmath$q$},
-{\rm i} \Omega_{n}))/2$ and 
$K_{-} = (K(\mbox{\boldmath$q$},{\rm i} \Omega_{n})-K(-\mbox{\boldmath$q$},
-{\rm i} \Omega_{n}))/2$.
 We omitted the indices $\mbox{\boldmath$q$}$ and $\Omega_{n}$.
 The condition $ 1/|g| - K(\mbox{\boldmath$0$},0) + L(\mbox{\boldmath$0$},0) 
= 0 $ is equivalent to the gap equation and is realized in the superconducting 
state. This fact indicates the existence of the gap-less phase mode 
 which corresponds to the Bogoliubov-Anderson mode 
at the zero temperature.~\cite{rf:anderson} However, the phase mode have a 
dissipation due to the quasi-particle excitations at the finite temperature. 

 In the particle-hole symmetric case, 
$ K(\mbox{\boldmath$q$},{\rm i} \Omega_{n}) =
K(-\mbox{\boldmath$q$},-{\rm i} \Omega_{n}) $ is satisfied, and 
the off-diagonal component $K_{-}$ vanishes. Therefore, the phase mode and the 
amplitude mode are completely decoupled. 
 However, High-$T_{{\rm c}}$ cuprates are strongly particle-hole asymmetric 
systems, as we have emphasized. 
 Therefore, the phase and amplitude fluctuations couples with each other 
through the off-diagonal component. 
 Anyway, since the off-diagonal component vanishes at $ \mbox{\boldmath$q$} = 
{\rm i} \Omega_{n} = 0 $, 
the T-matrix has a pole at $ \mbox{\boldmath$q$} = {\rm i} \Omega_{n} = 0 $ in 
the superconducting state (This is the Thouless criterion for $T_{{\rm c}}$). 
 Thus, the T-matrix in the superconducting state 
include both the phase and amplitude modes.

 We consider the effects of the superconducting fluctuations on the 
electronic state which are the origin of the pseudogap. 
 In the superconducting state the effects are mainly from the phase mode.  
However, the effects are small compared to those in the normal state. 
 Although the self-energy correction makes the spectrum broad, 
the effects make no significant difference in the low energy properties 
because the large superconducting gap rapidly grows below $T_{{\rm c}}$. 
(See Fig. 4 in the next section.)

 After carrying out the analytic continuation for the above expressions 
eqs.(2.9-16), we self-consistently determine the self-energy 
${\mit{\it \Sigma}}^{{\rm R}} (\mbox{\boldmath$k$}, \omega)$,
the order parameter $ \Delta_{\mbox{{\scriptsize \boldmath$k$}}} $,
normal and anomalous Green functions 
$ {\mit{\it G}}^{{\rm R}} (\mbox{\boldmath$k$},\omega) $,
$ {\mit{\it F}}^{{\rm R}} (\mbox{\boldmath$k$},\omega) $ 
and $ 2 \times 2 $ T-matrix on the real frequency. The calculation is carried 
out both in the normal state and in the superconducting state. 

 The effects of the superconducting fluctuations are included in the 
self-energy ${\mit{\it \Sigma}}^{{\rm R}} (\mbox{\boldmath$k$}, \omega)$. 
By neglecting the self-energy, our formalism reduces to the BCS mean field 
theory. 
 In the normal state, off-diagonal components 
$ \Delta_{\mbox{{\scriptsize \boldmath$k$}}} $,
$ {\mit{\it F}} (\mbox{\boldmath$k$},{\rm i} \omega_{n}) $ and 
$ L (\mbox{\boldmath$q$},{\rm i} \Omega_{n})$ vanish. 
 In this case, it can be easily confirmed that the above set of equations 
arrives at the self-consistent T-matrix calculation used in the normal state. 
 The self-consistent T-matrix calculation gives a unified description for 
the pseudogap state, the superconducting state and their phase transition, 
although it is not precise on the critical point.  
 As we have done in ref. 20, we keep the 
$\alpha = 1 + g K(\mbox{\boldmath$0$},0) - g L(\mbox{\boldmath$0$},0)$ 
as small value $\alpha = 0.01$ in the superconducting state 
in order to avoid the singularity from the two-dimensionality. 
 This operation is justified in the quasi-two dimensional systems which 
High-$T_{{\rm c}}$ cuprates are considered to be, because the weak 
three dimensionality surely remove the singularity. 
 The finite critical temperature is obtained by this operation. 
 The critical temperature $T_{{\rm c}}$ is more reduced as $\alpha$ 
is decreased. 
 That is a natural result because the value  $\alpha$ represents the 
three-dimensionality of the systems. 
 The choice of the value $\alpha$ makes no qualitative difference 
on the calculated results in this paper.

\section{Order Parameter and Single Particle Properties}

 Hereafter, we show the results of the self-consistent 
calculations for the set of equations in \S2. 
 In the main part of this paper, we choose the coupling constant $g = -1.0$, 
or $g = -2.0$. 
 Both cases give the qualitatively same results. 
 In case of $g = -1.0$, mean field critical temperature $T_{{\rm MF}} = 0.194$,
and the calculated critical temperature $T_{{\rm c}} = 0.099$.
 In case of $g = -2.0$, $T_{{\rm MF}} = 0.472$, and $T_{{\rm c}} = 0.212$. 
 The obtained phase diagram is shown in Fig. 3, which is similar 
to that in ref. 20. 
 Here, we have determined the critical temperature $T_{{\rm c}}$ as the 
highest temperature where the solution for the superconducting state is 
obtained.

\begin{figure}[htbp]
  \begin{center}
   \epsfysize=6cm
$$\epsffile{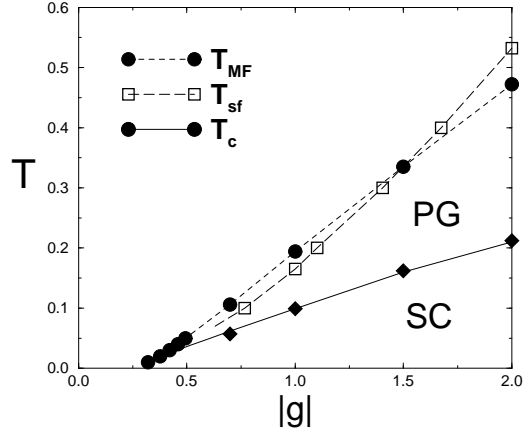}$$
%   \epsfile{file=phasediagram.eps,height=6cm}
    \caption{The obtained phase diagram. 
              The closed circles show the critical 
              temperature based on the mean field theory ($T_{{\rm MF}}$). 
              The closed diamonds show the critical 
              temperature ($T_{{\rm c}}$) suppressed by the fluctuations. 
              The suppression becomes remarkable in the strong coupling 
              region. 
              The open squares correspond to the temperature $T_{{\rm sf}}$ 
              where $1/|g| - \chi_{0}(\mbox{\boldmath$0$},0) = 0.1$. 
       }    
  \end{center}
\end{figure}

 The suppression of $T_{{\rm c}}$ from $T_{{\rm MF}}$ becomes remarkable 
with increasing the coupling constant $|g|$. 
 In the strong coupling case, the pseudogap state appears in the wide 
temperature region. 
 We show the region where 
$0 \le 1/|g| - \chi_{0}(\mbox{\boldmath$0$},0) \le 0.1$ in Fig. 3. 
 In case of the Gaussian fluctuation, 
$1/|g| - \chi_{0}(\mbox{\boldmath$0$},0) \cong 
\rho_{{\rm d}}(0) \frac{T-T_{{\rm c}}}{T_{{\rm c}}}$. Therefore, the width of 
the region is scaled by $T_{{\rm c}}$. 
 Our result shows that the region is enlarged in the strong coupling case. 
This indicates the wide critical region and the wide pseudogap region. 

 Once the superconducting order occurs, the effects of the fluctuations are 
drastically suppressed. The main reason is the following two points. 
 The amplitude mode is suppressed owing to the growth of the order parameter. 
 Moreover, the weight of the phase mode shifts to high frequency because 
the dissipation is reduced in the ordered state.  
 As a result, the order parameter 
$ \Delta_{\mbox{{\scriptsize \boldmath$k$}}} = 
\Delta \varphi_{\mbox{{\scriptsize \boldmath$k$}}} $ grows more rapidly than 
the result of the BCS theory. 
The temperature dependence of the order parameter is shown in Fig. 4. 
The rapid growth of the order parameter is a common feature of the theories 
including the critical fluctuations. 
  It should be noticed that the dissipation of the two modes 
is reduced at the low temperature in accordance with the power law, 
${\rm Im} K(\mbox{\boldmath$q$},\Omega) \pm  
{\rm Im} L(\mbox{\boldmath$q$},\Omega) \propto \Omega^{4}$, 
while it is exponentially reduced in the $s$-wave superconductor. 
 The power law is due to the gap node. Thus, the dissipation remains 
in the $d$-wave case more than in the $s$-wave case.

\begin{figure}[htbp]
  \begin{center}
   \epsfysize=6cm
$$\epsffile{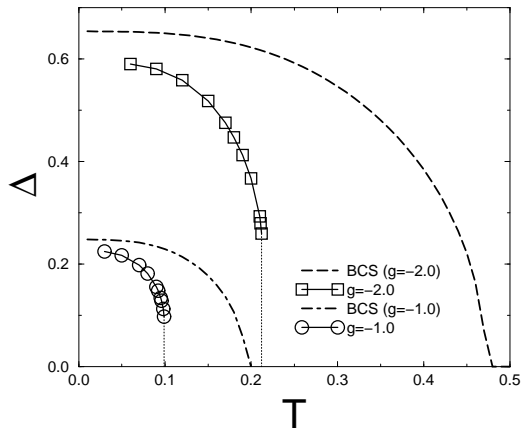}$$
%   \epsfile{file=Delta.eps,height=6cm}
    \caption{The growth of the order parameter $\Delta$. 
       The open circles and the open squares are the calculated results 
       for $g=-1.0$ and $g=-2.0$, respectively. 
       The dashed-dotted and long-dashed lines show the results of the 
       $d$-wave BCS theory for $g=-1.0$ and $g=-2.0$, respectively.
       The order parameter grows more rapidly than the BCS result.
       }    
  \end{center}
\end{figure}

 The rapid growth of the order parameter should be seen in various probes. 
 For example, the London penetration depth is a typical one (see \S5). 
 However, the rapid growth is a more general feature of the superconductivity 
caused by the electron correlation. 
 For example, the rapid growth of the order parameter is also shown 
in the FLEX calculation.~\cite{rf:takimoto} 
 We consider that the rapid growth obtained by the FLEX calculation is caused 
by the suppression of the de-pairing effect due to the low frequency spin 
fluctuations. 
 This effect is different from that calculated in this paper. 
 Both effects exist in High-$T_{{\rm c}}$ cuprates. 
 Since the de-pairing effect is suppressed by the pseudogap, 
it is expected that the effect shown in this paper becomes dominant 
with under-doping.

 Next, we show the results for the single particle properties. 
 The single particle spectral weight 
$A(\mbox{\boldmath$k$},\omega) = - 
\frac{1}{\pi} {\rm Im} {\mit{\it G}}^{{\rm R}} (\mbox{\boldmath$k$}, \omega)$  
is shown in Fig. 5. 
 The pseudogap appears in the normal state $T \geq T_{{\rm c}}$. 
 Since the calculation using the TDGL expansion is inappropriate 
in the high frequency region, the structure at high frequency is different 
from the result in ref. 20. 
 However, the broad and asymmetric structure of the pseudogap at the low 
frequency is obtained similarly. 
 As is pointed out in ref. 20, the asymmetric structure is 
essential for the self-consistent solution showing the pseudogap. 
 The strong particle-hole asymmetry originally exists in High-$T_{{\rm c}}$ 
cuprates and stabilizes the self-consistent solution showing the pseudogap 
state. 
 On the other hand, the gap structure becomes clear and symmetric 
below $ T_{{\rm c}} $. 
 This is a usual feature of the superconducting gap. 
 This change is smooth, but rapid because of the rapid growth of the 
order parameter. 
 The pseudogap is caused by the self-energy correction due to the 
superconducting fluctuations, while the superconducting gap is caused by the 
superconducting order. 
 The self-energy correction are reduced in the superconducting state 
because the fluctuations are suppressed. 

\begin{figure}[htbp]
  \begin{center}
   \epsfysize=6cm
$$\epsffile{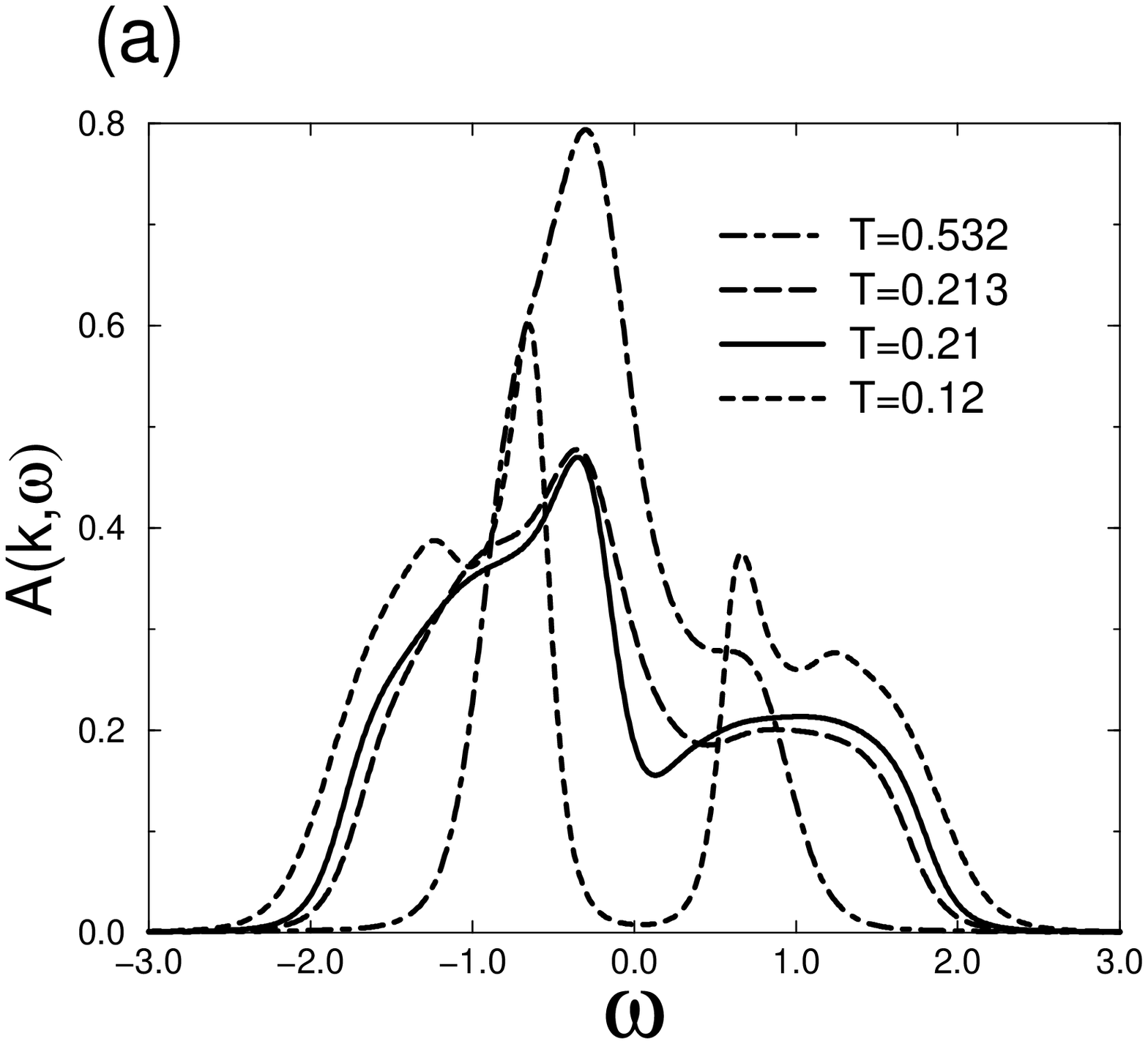}
%   \epsfile{file=spectr.eps,height=6cm}
\hspace{10mm}
   \epsfysize=6cm
\epsffile{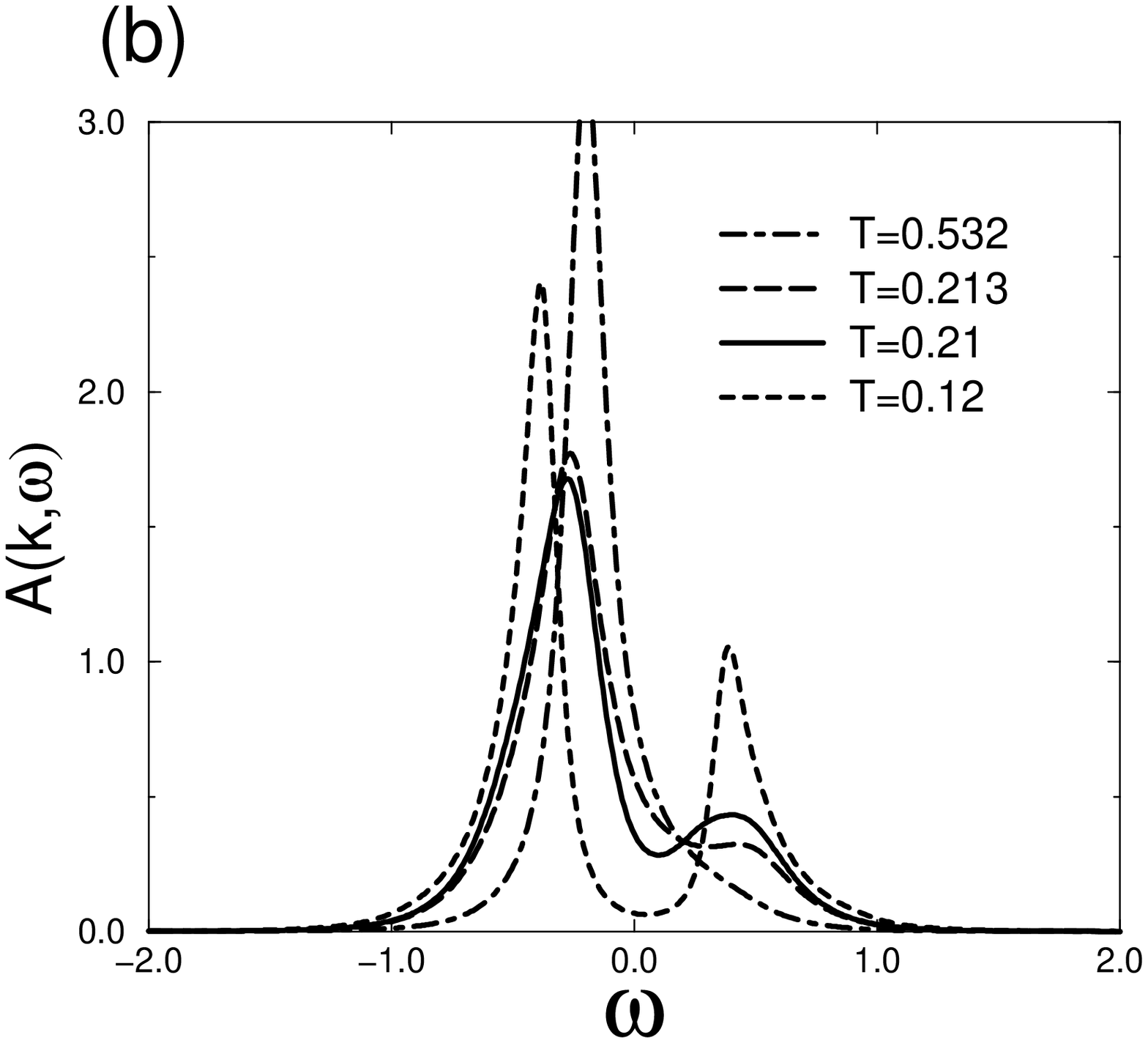}$$
%   \epsfile{file=spectrcold.eps,height=6cm}
    \caption{The single particle spectral weight at 
             (a) $\mbox{\boldmath$k$}=(\pi,0.15 \pi)$, 
             (b) $\mbox{\boldmath$k$}=(0.5 \pi,0.25 \pi)$ ,
             just bellow the Fermi level.
             The dash-dotted ($T=0.532$) and long-dashed ($T=0.213$) lines are 
             the results above $T_{{\rm c}}$. $T=0.213$ is just above 
             $T_{{\rm c}}$ and corresponds to the pseudogap state. 
             The solid ($T=0.21$) and the dashed lines ($T=0.12$) are results 
             below $T_{{\rm c}}$. 
             The pseudogap smoothly changes to the superconducting gap. 
             The gap structure becomes sharp and symmetric 
             in the superconducting state. 
             }
  \end{center}
\end{figure}

 It should be noticed that the energy scale of the pseudogap and that of the 
superconducting gap are similar including their momentum dependence. 
  The broad pseudogap, the sharp superconducting gap and the same energy scale 
are the important properties observed by ARPES.~\cite{rf:ARPES} 
 However, ARPES experiments can not refer to the asymmetric or symmetric 
structure, since ARPES measures only the spectrum below the Fermi energy. 
 Moreover, the asymmetric structure disappears by summing up the momentum 
corresponding to the experimental resolving power. 
 The details for the single particle spectral weight are shown in the separate 
paper.~\cite{rf:jujoyanase}

 Here, we briefly comment on the properties of the T-matrix 
in the pseudogap state. 
 The calculation in this paper justifies the calculation using 
the TDGL expansion in the pseudogap state~\cite{rf:yanasePG}. 
 Our calculation explicitly shows 
the properties of the TDGL parameters reviewed in \S2. 
 The effects of the renormalization are confirmed. 
 Both $a_{2}$ and $b$ are reduced in the pseudogap state. 
 In the TDGL expansion, we have neglected the quadratic term in $ \Omega $ 
which is emphasized as the lowest order term in the particle-hole 
symmetric case.~\cite{rf:tchernyshyov}  
 However, the quadratic term can be neglected near $T_{{\rm c}}$ 
because it is a higher order term than the linear term which is present here. 
 Moreover, our calculation shows that the quadratic term is reduced by the 
renormalization effect. 
 The coefficient of the quadratic term $a_{3}$ is proportional to 
the parameter $b$, $a_{3} =(2/\bar{v}^{2}) b $ in the Gaussian fluctuation. 
 Here, $\bar{v}$ is the mean value of the quasi-particle velocity on the 
Fermi surface. 
 The coefficient $a_{3}$ are reduced by the pseudogap with the parameter $b$. 
 Therefore, the quadratic term is not important in the pseudogap state. 
 On the other hand, 
 The quadratic term is restored in the superconducting state, 
although the sign is opposite to that above $T_{{\rm c}}$. 
 Therefore, the T-matrix becomes symmetric in the superconducting state. 
 That is a natural result because the superconductivity mixes the particles 
and the holes.

 We can see more clearly the character of the phase transition 
from the pseudogap state to the superconducting state 
by showing the density of states (DOS) in Fig. 6.  
 The DOS $\rho(\omega) = \sum_{\mbox{\boldmath$k$}} 
A(\mbox{\boldmath$k$},\omega) $ shows the broad pseudogap in the normal state. 
 The DOS remains near the Fermi level to some extent. 
 These features reflect the broad and asymmetric structure of the single 
particle spectral weight $A(\mbox{\boldmath$k$},\omega)$. 
 The pseudogap grows with approaching the critical point. 
 The superconducting phase transition takes place by the remained DOS. 
 Once the superconducting order occurs, the gap becomes sharp with 
the rapid growth of the order parameter. 
 The change is rapid but smooth across $ T_{{\rm c}} $. 
The DOS shows the linear energy dependence 
$\rho(\omega) \propto \omega$ at the low temperature. This is a characteristic
feature of the $d_{x^{2}-y^{2}}$-wave superconductivity.

\begin{figure}[htbp]
  \begin{center}
   \epsfysize=6cm
$$\epsffile{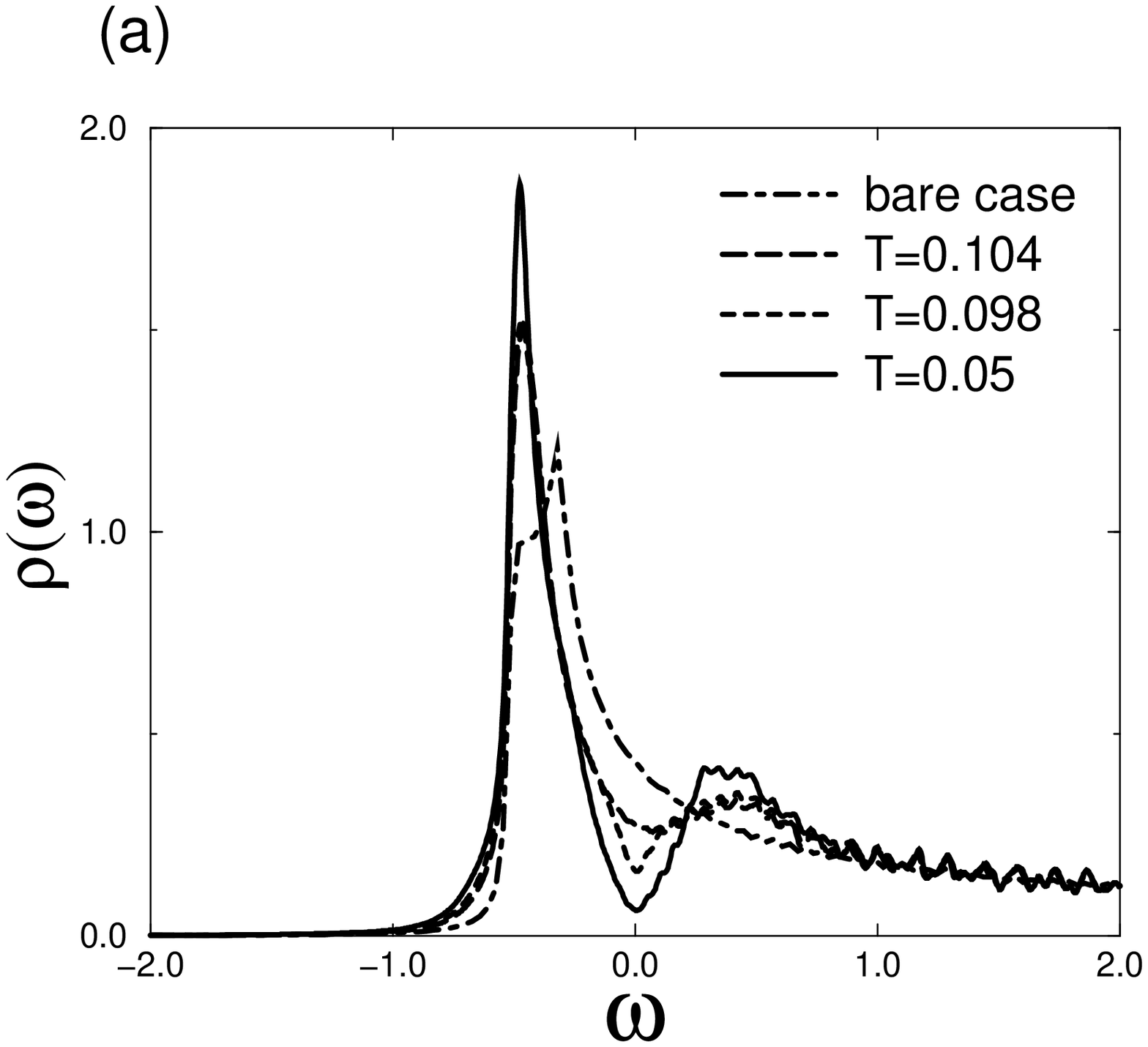}
%   \epsfile{file=DOSsmall.eps,height=6cm}
\hspace{10mm}
   \epsfysize=6cm
\epsffile{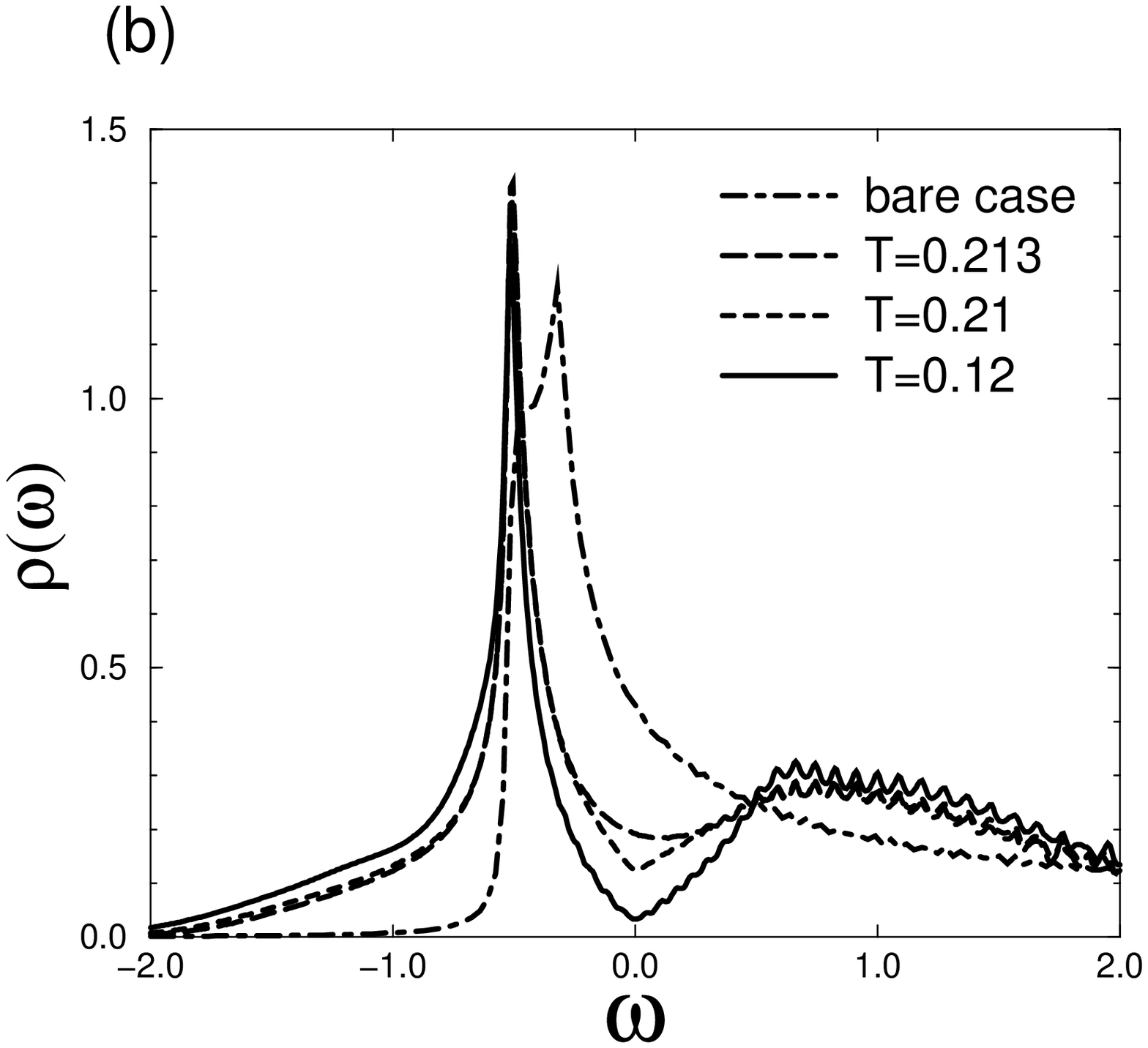}$$
%   \epsfile{file=DOS.eps,height=6cm}
    \caption{The density of states for (a) $g=-1.0$ and (b) $g=-2.0$. 
             The dash-dotted lines show the non-interacting DOS. 
             The long-dashed lines ((a) $T=0.104$ (b) $T=0.213$) show 
             the pseudogap state above $T_{{\rm c}}$. 
             The dashed ((a) $T=0.098$ (b) $T=0.21$) and solid 
             ((a) $T=0.05$ (b)$T=0.12$) lines are the results 
             bellow $T_{{\rm c}}$. 
             It should be noticed that the gap scale is almost the same 
             between in the pseudogap state and in the superconducting state. 
             } 
  \end{center}
\end{figure}

 It should be noticed that the energy scale of the pseudogap and that of the 
superconducting gap are almost the same. 
 When the coupling constant $|g|$ increases, the energy scale of the pseudogap 
increases with that of the superconducting gap. 
 The energy scale is almost independent of the temperature. 
 These features are consistent with the important results of the spectroscopic 
measurements.\cite{rf:ARPES,rf:renner} 
 The same gap scale is not so obvious in the resonance scattering scenario, 
while it is obvious in the NSR scenario. 
 However, the same gap scale is an expected result because the two phenomena, 
the pseudogap and the superconductivity, have the same origin in our scenario. 
 This is actually confirmed by the above results.

\section{Magnetic Properties}

 In this section, we show the results for the magnetic properties. 
 We calculate the dynamical spin susceptibility 
$\chi_{{\rm s}}(\mbox{\boldmath$k$}, \omega)$ by using the random phase 
approximation (RPA). 
 The exchange enhancement is taken into account within the RPA. 
\begin{eqnarray}
  \chi_{{\rm s}}^{{\rm R}}(\mbox{\boldmath$q$}, \omega) & = & 
  \frac{\chi_{0}^{{\rm R}}(\mbox{\boldmath$q$}, \omega)}
       {1 - U \chi_{0}^{{\rm R}}(\mbox{\boldmath$q$}, \omega)},
\\
   \chi_{0}(\mbox{\boldmath$q$}, {\rm i} \omega_{n}) & = & 
   -T \sum_{\mbox{\boldmath$k$},\omega_{m}} 
     [{\mit{\it G}} (\mbox{\boldmath$k$},{\rm i} \omega_{m}) 
      {\mit{\it G}} (\mbox{\boldmath$k$}+\mbox{\boldmath$q$},
                  {\rm i} \Omega_{m} + {\rm i} \omega_{n}) 
\nonumber \\
     & &  + {\mit{\it F}} (\mbox{\boldmath$k$},{\rm i} \omega_{m}) 
         {\mit{\it F}} (\mbox{\boldmath$k$}+\mbox{\boldmath$q$},
                  {\rm i} \Omega_{m} + {\rm i} \omega_{n})].
\end{eqnarray}

 We fix the enhancement parameter $U=1.5$, afterward. 
The following results are not affected by the choice of the parameter, 
qualitatively.

 Generally speaking, the AL term and the MT term is considered 
as corrections by the fluctuations on the two-body correlation function 
above $T_{{\rm c}}$.~\cite{rf:AL,rf:MT} 
 However, the AL term dose not exist in calculating the spin susceptibility 
$ \chi_{{\rm s}}^{{\rm R}} (\mbox{\boldmath$q$}, \omega) $. 
 This fact is understood by considering the spin index for the spin 
singlet pairing. 
 The contribution from the MT term is small in case of the d-wave pairing, and 
suppressed by the slight elastic scattering.~\cite{rf:eschrig}  
 Therefore, we have only to calculate the effects through the single particle 
properties (that is, pseudogap) as effects of the superconducting 
fluctuations.

 The NMR spin-lattice relaxation rate $1/T_{1}$ and spin-echo decay rate 
$1/T_{2G}$ are calculated by the following expressions. 
\begin{eqnarray}
  1/T_{1}T & = & \sum_{\mbox{\boldmath$q$}} F_{\perp}(\mbox{\boldmath$q$})  
             [\frac{1}{\omega} 
             {\rm Im} \chi_{{\rm s}}^{{\rm R}} (\mbox{\boldmath$q$}, \omega) 
             \mid_{\omega \to 0}],
\\
  1/T_{2G}^{2} & = & \sum_{\mbox{\boldmath$q$}} 
                     [F_{\parallel}(\mbox{\boldmath$q$}) 
        {\rm Re} \chi_{{\rm s}}^{{\rm R}} (\mbox{\boldmath$q$}, 0)]^{2}
                   - [\sum_{\mbox{\boldmath$q$}} 
                      F_{\parallel}(\mbox{\boldmath$q$}) 
        {\rm Re} \chi_{{\rm s}}^{{\rm R}} (\mbox{\boldmath$q$}, 0)]^{2}.
\end{eqnarray}

 Here, $F_{\perp}(\mbox{\boldmath$q$}) = \frac{1}{2} [\{A_{1} + 2 B 
(\cos q_{x}+\cos q_{y})\}^{2} + \{A_{2} + 2 B (\cos q_{x}+\cos q_{y})\}^{2}] $
 and $F_{\parallel}(\mbox{\boldmath$q$}) = 
\{A_{2} + 2 B (\cos q_{x}+\cos q_{y})\}^{2}$.
 The hyperfine coupling constants $A_{1}, A_{2}$ and $B$ are evaluated as 
$A_{1} = 0.84 B$ and $A_{2} = -4 B$.~\cite{rf:barzykin}

\begin{figure}[htbp]
  \begin{center}
   \epsfysize=6cm
$$\epsffile{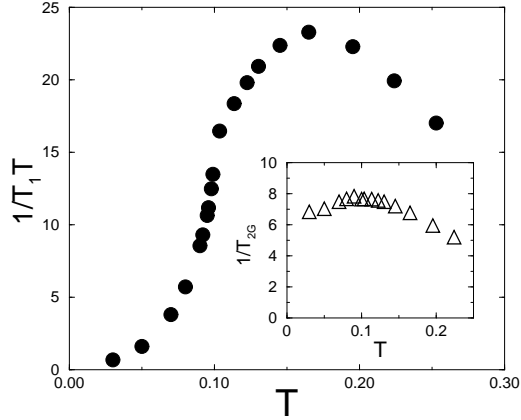}$$
%   \epsfile{file=T2Ginv.eps,height=6cm}
    \caption{The results for $1/T_{1}T$. 
             Here, $g=-1.0$. 
             The peak appears above $T_{{\rm c}}$. 
             $1/T_{1}T$ is reduced by the pseudogap and decreases with 
             approaching the critical point. 
             The inset shows the result for $1/T_{2G}$. 
             }
  \end{center}
\end{figure}

 The calculated results are shown in Fig. 7. 
 In the normal state far above $T_{{\rm c}}$, $1/T_{1}T$ increases with 
lowering the temperature owing to the exchange enhancement. 
 It shows a peak above $T_{{\rm c}}$ and decreases with lowering the 
temperature below $T^{*}$. 
The decrease is an effect of the superconducting fluctuations. 
 It is the well-known pseudogap phenomena in NMR $1/T_{1}T$.~\cite{rf:NMR} 
 The imaginary part of the spin susceptibility at the low frequency 
reflects the DOS. 
 The superconducting fluctuations give rise to the pseudogap in the DOS, and 
reduce the weight of the spin fluctuations at the low frequency. 
 Thus, the pseudogap observed in NMR $1/T_{1}T$ takes place through the 
single particle properties. 
 However, the decrease in the pseudogap state is 
rather moderate than in the superconducting state. 
 With the growth of the sharp superconducting gap, 
$1/T_{1}T$ decreases rapidly. 
 The BCS-like behaviors are obtained in the superconducting state
since the large superconducting gap opens rapidly. 
 These features are consistent with the experimental results. 

 On the other hand, NMR $1/T_{2G}$ shows rather weak temperature dependence 
both in the pseudogap state and in the superconducting state 
(see the inset of Fig. 7). 
 NMR $1/T_{2G}$ increases with decreasing temperature in the normal state 
far above $T_{{\rm c}}$ owing to the exchange enhancement. 
 The pseudogap also affects the NMR $1/T_{2G}$. 
 The increase of $1/T_{2G}$ becomes moderate in the pseudogap state. 
 Near $T_{{\rm c}}$, NMR $1/T_{2G}$ is almost independent of the temperature. 
 These are the effects of the pseudogap. 
 However, the effects on $1/T_{2G}$ are weaker than those on $1/T_{1}T$. 
 In the superconducting state, $1/T_{2G}$ decreases moderately with decreasing 
the temperature as a characteristic result of the $d$-wave pairing.

 These results indicate that the effects of the pseudogap are weak on 
the real part of the spin susceptibility rather than on the imaginary part. 
 The dissipation (imaginary part) directly reflects the DOS at the low energy. 
 However, the static properties (real part) do not necessarily so.  
 In other wards, the pseudogap suppresses the weight of the spin fluctuations 
only at the low frequency. 
 Since $1/T_{2G}$ reflects the total weight of the spin fluctuations, 
the effect of the pseudogap is weak on $1/T_{2G}$. 
 In particular, ${\rm Re} \chi_{{\rm s}}^{{\rm R}} (\mbox{\boldmath$q$}, 0)$ 
around $\mbox{\boldmath$q$}=\mbox{\boldmath$Q$}=(\pi,\pi)$ 
is not reduced so much by the $d$-wave pseudogap, while 
${\rm Re} \chi_{{\rm s}}^{{\rm R}} (\mbox{\boldmath$q$}, 0)$ is remarkably 
reduced at $\mbox{\boldmath$q$}=(0,0)$. 
 The momentum dependence of the hyperfine coupling  
$F_{\parallel}(\mbox{\boldmath$q$})$ further weaken the effects of the 
pseudogap on $1/T_{2G}$. 
 The above features are in common with those in the superconducting 
state.~\cite{rf:bulut} 
 That is a natural result because the pseudogap and the superconducting gap 
have the same $d_{x^{2}-y^{2}}$-wave form. 
 In other words, it is natural that the scaling law for the spin fluctuations 
is violated by the pseudogap, because it is violated 
in the superconducting state. 
 The above results are qualitatively consistent with the features indicated by 
the NMR experiments.~\cite{rf:NMR,rf:tokunaga}
 Minutely speaking, the different behaviors of $1/T_{2G}$ have been 
reported for different High-$T_{{\rm c}}$ 
compounds.~\cite{rf:NMR,rf:tokunaga,rf:goto} 
 There is an idea that attributes the difference to the effects of 
the interlayer coupling.~\cite{rf:goto} 
 Anyway, the relatively weak effect of the pseudogap 
on $1/T_{2G}$ than on $1/T_{1}T$
is observed in common. The qualitatively consistent behaviors in the above 
sense are obtained here.

 The remaining part of this section is concerned with the neutron resonance 
peak observed by the neutron scattering 
experiments.~\cite{rf:neutrons,rf:neutronp} 
 The neutron resonance peak is observed in the superconducting state from 
optimally-doped to under-doped region. 
 The recent measurements have found the development of the weak resonance peak 
in the pseudogap state.~\cite{rf:neutronp} 
 The resonance peak in the superconducting state is attributed to the 
anti-ferromagnetic correlation and the $d$-wave 
superconductivity.~\cite{rf:takimoto,rf:yoshikawa,rf:morr} 
 Our result shows the sharp resonance peak in the superconducting state, 
which is caused by the same mechanism.
 Moreover, we investigate the behavior of the resonance peak in the pseudogap 
state. 
 The neutron scattering intensity is proportional to the imaginary part of the 
spin susceptibility, 
$ {\rm Im} \chi_{{\rm s}}^{{\rm R}} (\mbox{\boldmath$q$}, \omega) $. 
 We show the calculated results for 
$ {\rm Im} \chi_{{\rm s}}^{{\rm R}} (\mbox{\boldmath$q$}, \omega) $  
at $\mbox{\boldmath$q$}=\mbox{\boldmath$Q$}$ in Fig. 8. 
 Here, $ {\rm Im} \chi_{{\rm s}}^{{\rm R}} (\mbox{\boldmath$Q$}, \omega) $ 
is expressed as follows, 

\begin{eqnarray}
  {\rm Im} \chi_{{\rm s}}^{{\rm R}} (\mbox{\boldmath$Q$}, \omega) = 
    \frac{{\rm Im} \chi_{0}^{{\rm R}} (\mbox{\boldmath$Q$}, \omega)}
       {[1 - U {\rm Re} \chi_{0}^{{\rm R}}(\mbox{\boldmath$Q$}, \omega)]^{2}
        + [U {\rm Im} \chi_{0}^{{\rm R}}(\mbox{\boldmath$Q$}, \omega)]^{2}}.
\end{eqnarray}

\begin{figure}[htbp]
  \begin{center}
   \epsfysize=6cm
$$\epsffile{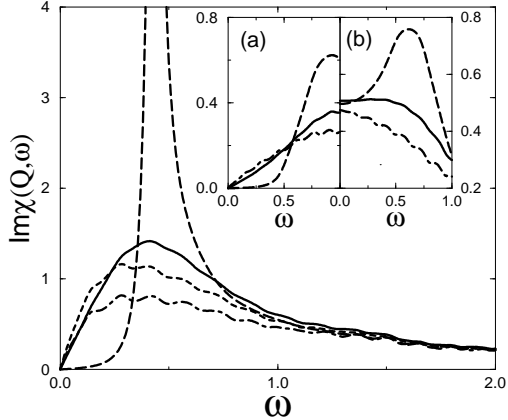}$$
%   \epsfile{file=imKaiQRPA.eps,height=6cm}
    \caption{The results for $ {\rm Im} \chi_{{\rm s}}^{{\rm R}} 
             (\mbox{\boldmath$Q$}, \omega) $. Here, $g=-1.0$. 
             The inset shows 
             (a) $ {\rm Im} \chi_{{\rm 0}}^{{\rm R}} 
             (\mbox{\boldmath$Q$}, \omega) $ and 
             (b) $ {\rm Re} \chi_{{\rm 0}}^{{\rm R}} 
             (\mbox{\boldmath$Q$}, \omega) $, respectively. 
             The long-dashed lines, the solid lines, the dashed-lines 
             and the dash-dotted lines 
             correspond to $T=0.050$ (superconducting state), 
             $T=0.104$ (pseudogap state), $T=0.165$ and $T=0.253$, 
             respectively.  
             }
  \end{center}
\end{figure}

 In the superconducting state, 
${\rm Re} \chi_{0}^{{\rm R}}(\mbox{\boldmath$Q$}, \omega)$ increases with 
$\omega$ and ${\rm Im} \chi_{0}^{{\rm R}}(\mbox{\boldmath$Q$}, \omega)$ 
is suppressed at the low frequency $\omega < 2 \Delta_{{\rm max}}$. 
 Here, $\Delta_{{\rm max}}$ is the maximum gap energy. 
 Therefore, the denominator of eq.(4.5) is remarkably small at the finite 
frequency $\omega_{{\rm r}} < 2 \Delta_{{\rm max}}$ 
when the system is near the anti-ferromagnetic instability, 
$1 - U {\rm Re} \chi_{0}^{{\rm R}}(\mbox{\boldmath$Q$}, 0) \ll 1$. 
 Then, the sharp resonance peak appears at $\omega=\omega_{{\rm r}}$. 
 Our result shows the same features and the sharp resonance peak in the 
superconducting state. 
 In the pseudogap state, $\chi_{0}$ behaves similarly. 
${\rm Re} \chi_{0}^{{\rm R}}(\mbox{\boldmath$Q$}, \omega)$ has the positive 
slope with respect to $\omega$. The dissipation 
${\rm Im} \chi_{0}^{{\rm R}}(\mbox{\boldmath$Q$}, \omega)$ is reduced 
at the low frequency. 
 These behaviors are essential results of the pseudogap as a precursor of the 
superconductivity. 
 However, the features are rather weak compared to the superconducting state. 
In particular, the finite dissipation remains in the pseudogap state 
because of the finite DOS at the Fermi energy. 
 As a result, the peak develops as a precursor of the resonance peak, 
although it is weak and broad. 
 The peak appears at the slightly lower frequency than 
$\omega_{{\rm r}}$. The peak smoothly develops to the sharp peak in the 
superconducting state. 
 Far above $T_{{\rm c}}$, the conventional behaviors of the normal state
is obtained. 
 We can see that the low frequency component of the spin fluctuations is 
suppressed by the pseudogap, however, the weight is transfered to the high 
frequency. These behaviors are not obtained by considering 
the spin fluctuations alone. 
 The transfered high frequency component is effective as 
the pairing interaction, and contribute to NMR $1/T_{2G}$.

\section{Transport Properties}

 In this section, we show the results for the transport properties. 
 We investigate the effects of the pseudogap and the superconductivity 
on the transport phenomena. 
 The experimental results clearly show the pseudogap phenomena in 
the {\it c}-axis transport.~\cite{rf:homes,rf:transport} 
 On the other hand, the in-plane transport does not clearly show 
the pseudogap. 
 The different effects of the pseudogap on the in-plane and {\it c}-axis 
transport are understood in the following way. 
 As is previously emphasized,~\cite{rf:yanaseTR,rf:ioffe} the momentum 
dependence of the interlayer hopping matrix element 
$t_{\perp}(\mbox{\boldmath$k$})$ plays an important role in the {\it c}-axis 
transport. 
 The transfer matrix $t_{\perp}(\mbox{\boldmath$k$})$ obtained 
by the band calculation is expressed as,~\cite{rf:okanderson}   
\begin{eqnarray}
  \label{eq:c-hopping}
  t_{\perp}(\mbox{\boldmath$k$}) =  
  t_{{\rm c}} (\frac{{\rm cos} k_{x} - {\rm cos} k_{y}}{2})^{2}. 
\end{eqnarray}
 Because of this momentum dependence, the quasiparticles near 
$\mbox{\boldmath$k$}=(\pi/2,\pi/2)$ ('cold spot') does not contribute to 
the {\it c}-axis transport. 
 The {\it c}-axis transport is mainly determined by the quasiparticles near 
$\mbox{\boldmath$k$}=(\pi,0)$ ('hot spot'), although the in-plane transport 
phenomena are dominated by the contribution from the 'cold spot'. 
 Therefore, the qualitatively different properties between the 
in-plane and {\it c}-axis transport are yielded. 
 Since the pseudogap and the superconducting gap are large at 'hot spot', 
the {\it c}-axis transport reflects the pseudogap 
more clearly than the in-plane transport.  
 The experimental results actually indicate so. 
 The {\it c}-axis resistivity increases with decreasing temperature 
in the pseudogap state, while the in-plane resistivity keeps the T-linear law 
and slightly deviates downward.~\cite{rf:transport}  
 The {\it c}-axis optical conductivity shows the gap structure in the 
pseudogap state, although the in-plane optical conductivity shows only the 
weak structure in the higher frequency region.~\cite{rf:homes}  
 The pseudogap observed in the {\it c}-axis optical conductivity smoothly  
changes to the superconducting gap.~\cite{rf:homes}  
 This fact also indicates the close relation between the pseudogap and the 
superconductivity. However, the pseudogap in the {\it c}-axis optical 
conductivity has never been explained on the basis of the pairing scenario.

 The anomalous properties 
above $T^{*}$ are well explained by considering the effects of the spin 
fluctuations.~\cite{rf:yanaseTR,rf:stojkovic,rf:kontani} 
 The scattering due to the anti-ferromagnetic spin-fluctuations is strong at 
'hot spot' and weak at 'cold spot'. 
 Since the 'hot spot' does not contribute to the in-plane conductivity even in 
the absence of the pseudogap,  
 the effects of the pseudogap on the in-plane transport are weak. 
 The detailed properties of the in-plane longitudinal transport in the 
pseudogap state are explained by considering the feedback effect of the 
pseudogap on the low frequency spin 
fluctuations.~\cite{rf:yanaseTR,rf:dahmtransport}

 Here, we calculate the conductivity by neglecting the vertex corrections. 
 Generally, the vertex correction due to the electron correlation is not 
important except for the factor arising from Umklapp 
scattering.~\cite{rf:yamada} 
 The conductivity $\sigma_{{\rm tot}}(\omega)$ is described by 
the normal fluid part and the superfluid part at the low frequency,  
$\sigma_{{\rm tot}}(\omega) = \sigma(\omega) 
 + {\rm i}/4 \pi \lambda^{2} (\omega + {\rm i} \delta) $. 
 The second term is proportional to the London constant 
(or the superfluid density), $\Lambda =1 /4 \pi \lambda^{2}$ 
and appears in the superconducting state, 
where $\lambda$ is the London penetration depth which we calculate later. 
 The in-plane and {\it c}-axis optical conductivity at the finite frequency 
is expressed by the normal fluid part in the following way, 
\begin{eqnarray}
  \sigma_{{\rm ab}}(\omega) & = & - \frac{e^{2}}{d} \frac{1}{\omega} 
     \sum_{\mbox{\boldmath$k$}} v^{2}(\mbox{\boldmath$k$})
     \int \frac{{\rm d} \omega'}{\pi} [f(\omega'+\omega)-f(\omega')]
\nonumber \\
 & & \times ({\rm Im} {\mit{\it G}}^{{\rm R}}(\mbox{\boldmath$k$},\omega')
      {\rm Im} {\mit{\it G}}^{{\rm R}}(\mbox{\boldmath$k$},\omega+\omega')
     +{\rm Im} {\mit{\it F}}^{{\rm R}}(\mbox{\boldmath$k$},\omega')
      {\rm Im} {\mit{\it F}}^{{\rm R}}(\mbox{\boldmath$k$},\omega+\omega')),
\\
  \sigma_{{\rm c}}(\omega) & = & -4 d e^{2} \frac{1}{\omega}  
     \sum_{\mbox{\boldmath$k$}} t_{\perp}^{2}(\mbox{\boldmath$k$})
     \int \frac{{\rm d} \omega'}{\pi} [f(\omega'+\omega)-f(\omega')]
\nonumber \\
 & & \times ({\rm Im} {\mit{\it G}}^{{\rm R}}(\mbox{\boldmath$k$},\omega')
      {\rm Im} {\mit{\it G}}^{{\rm R}}(\mbox{\boldmath$k$},\omega+\omega')
     +{\rm Im} {\mit{\it F}}^{{\rm R}}(\mbox{\boldmath$k$},\omega')
      {\rm Im} {\mit{\it F}}^{{\rm R}}(\mbox{\boldmath$k$},\omega+\omega')).
\end{eqnarray} 
 Here, $d$ is the interlayer distance, and 
$v(\mbox{\boldmath$k$}) = \sqrt{v_{x}^{2} + v_{y}^{2}}$ is the in-plane 
velocity, where 
$v_{\mu}=\frac{\partial \varepsilon_{\mbox{{\scriptsize \boldmath$k$}}}}
{\partial k_{\mu}}$. 
 The velocity $v(\mbox{\boldmath$k$})$ is almost independent of the momentum 
on the Fermi surface in our model. 
 We have neglected the weak $k_{z}$ dependence of the electronic state. 
 This corresponds to the lowest order calculation with respect to $t_{\rm c}$. 
 We have neglected the term proportional to the delta function 
$\delta(\omega)$ which comes from the superfluid part.

 Here, we comment on the vertex correction neglected here. 
 In particular, the AL term is considered to be important for the 
transport.~\cite{rf:AL} 
 We think that the AL term is not important except for the narrow region 
near $T_{{\rm c}}$. This expectation is explained as follows. 
 The current vertex parallel to the plane in the AL term 
$J_{ab}(\mbox{\boldmath$q$})$ is proportional to the TDGL parameter $b$,  
$J_{ab}(\mbox{\boldmath$q$}) \propto b \mbox{\boldmath$q$} $. 
 This corresponds to the velocity of the fluctuating Cooper pairs and 
is small in the strong coupling case. 
 Thus, the small $b$ indicates the small velocity of the fluctuating Cooper 
pairs as well as the strong fluctuations. 
 The factors $b$ from the current vertex and from the pair propagator 
cancel each other in two-dimension, 
$\sigma_{{\rm AL}} \propto 1/t_{0}$.~\cite{rf:AL} 
 On the other hand, the effects on the single particle properties are 
estimated as $1/\tau_{\mbox{{\scriptsize \boldmath$k$}}} = 
- {\rm Im} {\mit{\it \Sigma}}^{{\rm R}} (\mbox{\boldmath$k$}, \omega) \propto 
1/\sqrt{b t_{0}} $. 
 Since the singularity about $t_{0}$ is stronger in $\sigma_{{\rm AL}}$ than 
in $1/\tau_{\mbox{{\scriptsize \boldmath$k$}}}$, and since the parameter $b$ 
is large, the AL term is more important than the latter 
in the weak coupling limit. 
 However, the situation is quite different in the strong coupling case. 
The effects on the single particle properties can appear at $T^{*}$ before 
the AL term dominate the large conductivity from the `cold spot'. 
 On the other hand, the conductivity is small along the $c$-axis.  
However, we can neglect the AL term along the $c$-axis in the quasi-two 
dimensional systems because the AL term along the $c$-axis is 
a higher order term with respect to $t_{{\rm c}}/t$. 
 It is because the current vertex $J_{c}(\mbox{\boldmath$q$})$ is quadratic 
$J_{c}(\mbox{\boldmath$q$}) \propto t_{{\rm c}}^{2}$, 
although the quasi-particle velocity is linear 
$v_{{\rm c}} \propto t_{{\rm c}}$. 
 The MT term is considered to be suppressed by the elastic scattering 
in the $d$-wave case. 
 It has been pointed out that the vertex correction due to the 
spin fluctuations is important for the enhancement of 
the Hall coefficient.~\cite{rf:kontani} 
 However, the correction is shown to be not important for the longitudinal 
conductivity.~\cite{rf:kontani}

 Hereafter, we normalize the conductivity by the constant factor as  
$\sigma_{{\rm c}}(\omega)=\sigma_{{\rm c}}(\omega)/\sigma_{{\rm c}}^{0}$ and 
$\sigma_{{\rm ab}}(\omega)=\sigma_{{\rm ab}}(\omega)/\sigma_{{\rm ab}}^{0}$, 
where $\sigma_{{\rm c}}^{0}=d e^{2}$ and $\sigma_{{\rm ab}}^{0}=e^{2}/d$. 
 Hereafter, we fix the coupling constant $g=-2.0$.

\begin{figure}[htbp]
  \begin{center}
   \epsfysize=6cm
$$\epsffile{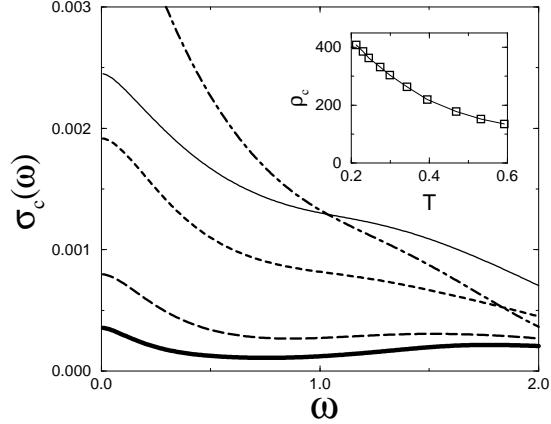}$$
%   \epsfile{file=SigmaC.eps,height=6cm}
    \caption{The c-axis optical conductivity due to the coherent process 
             for $T=0.395$ (dash-dotted line), $T=0.213$ (thin solid line), 
             $T=0.20$ (dashed line), $T=0.15$ (long-dashed line) and 
             $T=0.12$ (thick solid line). 
             The Drude peak is suppressed in the pseudogap state ($T=0.213$) 
             and the superconducting state ($T=0.20$,$T=0.15$,$T=0.12$). 
             The inset shows the normal state c-axis resistivity 
             $\rho_{{\rm c}}$. 
             }    
  \end{center}
\end{figure}

 The calculated results are shown in Figs. 9 and 10. We fix $t_{{\rm c}}=0.1$. 
First, the {\it c}-axis resistivity shows a semi-conductive behavior 
in the pseudogap state (the inset in Fig. 9). This is because the scattering 
due to the superconducting fluctuations increases with approaching the 
critical point. 
 Next, the results for the {\it c}-axis optical conductivity are shown 
in Fig. 9. 
 The Drude peak is remarkably suppressed by the pseudogap. 
 In other words, the coherent picture of the {\it c}-axis transport 
is broken by the pseudogap. 
 In the superconducting state, the {\it c}-axis optical conductivity is 
suppressed furthermore and shows the gap structure at low temperatures. 
 It should be noticed that the above results in the superconducting state are 
different from the expectation of the conventional $d$-wave BCS theory. 
 Since the gap node does not contribute to the {\it c}-axis transport, 
the {\it c}-axis transport behaves like a $s$-wave superconductor in the 
ordered state.

\begin{figure}[htbp]
  \begin{center}
   \epsfysize=6cm
$$\epsffile{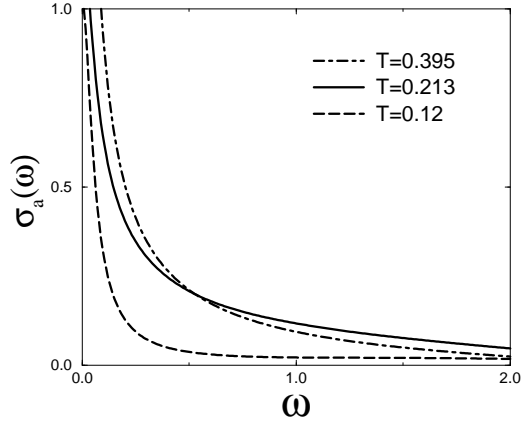}$$
%   \epsfile{file=SigmaA.eps,height=6cm}
    \caption{The in-plane optical conductivity in the normal state $T=0.395$ 
             (dash-dotted line), the pseudogap state $T=0.213$ (solid line) 
             and the superconducting state $T=0.12$ (long-dashed line), 
             respectively. 
             The Drude peak remains in all temperature range. 
             }
  \end{center}
\end{figure}

 The results for the in-plane optical conductivity are shown in Fig. 10.  
 The Drude peak remains even in the deeply superconducting state and 
the gap structure does not appear. 
 These features are characteristics of the $d$-wave symmetry and due to the 
contribution from the gap node. 
 The sharp Drude peak appears in the in-plane optical conductivity 
in the pseudogap state, owing to the contribution from the 
'cold spot' similarly. 
 The qualitatively different results for the {\it c}-axis transport 
originate in the momentum dependence of the interlayer hopping 
$t_{\perp}(\mbox{\boldmath$k$})$. 
 Thus, the coherent in-plane transport and the incoherent {\it c}-axis 
transport in the pseudogap state are understood simultaneously in a consistent 
way.

 As we can see from the above results, the coherent part for the {\it c}-axis 
transport is small in the pseudogap state. 
 Therefore, we can not neglect the incoherent process, especially in 
under-doped cuprates. The tunneling process is considered to be a typical 
incoherent process, although there are many other incoherent processes. 
 The contribution from the incoherent process have 
been discussed by Hirschfeld {\it et. al}~\cite{rf:incoherent}. 
 According to their formalism, we obtain the following expression 
for the contribution from the tunneling process, $\sigma_{{\rm inc}}(\omega)$. 
\begin{eqnarray}
  \sigma_{{\rm inc}}(\omega)  =  - d e^{2} t_{{\rm inc}}^{2} 
     \frac{1}{\omega}  
     \sum_{\mbox{\boldmath$k$},\mbox{\boldmath$k'$}} 
     \int \frac{{\rm d} \omega'}{\pi} [f(\omega'+\omega)-f(\omega')]
      {\rm Im} {\mit{\it G}}^{{\rm R}}(\mbox{\boldmath$k$},\omega')
      {\rm Im} {\mit{\it G}}^{{\rm R}}(\mbox{\boldmath$k'$},\omega+\omega'). 
\end{eqnarray}

 Here, we have neglected the momentum dependence of the tunneling matrix 
element $t_{{\rm inc}}$ for simplicity. 
 The above expression corresponds to the 
diffusive process in which the information for the in-plane momentum is lost. 
 The contribution from the anomalous Green function 
${\mit{\it F}}^{{\rm R}}(\mbox{\boldmath$k$},\omega)$ vanishes 
due to the $d$-wave symmetry.

\begin{figure}[htbp]
  \begin{center}
   \epsfysize=6cm
$$\epsffile{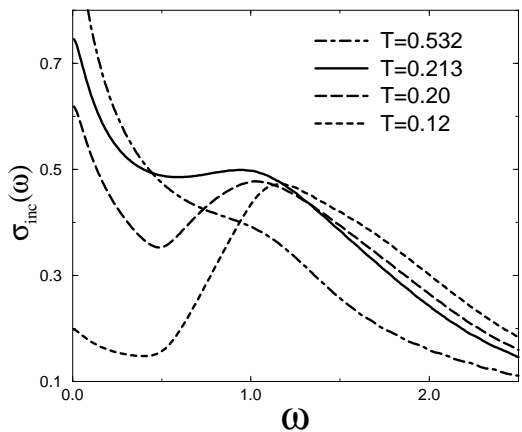}$$
%   \epsfile{file=SigmaI.eps,height=6cm}
    \caption{The c-axis optical conductivity due to the incoherent process 
             for $T=0.526$ (dash-dotted line), $T=0.213$ (solid line), 
             $T=0.20$ (long-dashed line), and $T=0.12$ (dashed line). 
             The weak gap structure is shown in the pseudogap state. 
             In the superconducting state,  
             The gap becomes deep with the same frequency scale. 
             }
  \end{center}
\end{figure}

 The results are shown in Fig. 11. We show the results normalized by the 
constant factor as 
$\sigma_{{\rm inc}}(\omega)=\sigma_{{\rm inc}}(\omega)/\sigma_{{\rm inc}}^{0}$,
 where $\sigma_{{\rm inc}}^{0}  = d e^{2} t_{{\rm inc}}^{2}$. 
Here, $\sigma_{{\rm inc}}(\omega)$ is suppressed at the low frequency by the 
pseudogap and shows the weak gap structure with the energy scale of 
$2 \Delta_{{\rm eff}}$. 
 Here, $\Delta_{{\rm eff}}$ is the energy scale of the pseudogap. 
 In the superconducting state, the gap structure becomes sharp, and 
the frequency scale of the gap does not change. 
 This is a natural result because the same energy scale between the pseudogap 
and the superconducting gap is shown in the DOS (Fig. 6). 
 These features are qualitatively consistent with the 
experimental results~\cite{rf:homes}. 
 Thus, our scenario for the pseudogap phenomena well explains the 
pseudogap observed in the {\it c}-axis optical conductivity and its 
smooth change to the superconducting gap. 
 Of course, the {\it dc}-conductivity from the incoherent process shows a 
semi-conductive behavior. 

 Recently, the {\it c}-axis optical conductivity has been discussed 
in connection with the violation of the f-sum rule.~\cite{rf:sumrule} 
 The f-sum rule is violated in the under-doped cuprates, although it is 
satisfied from the over- to optimally-doped region. 
 The violation has been attributed to the effects of the phase 
fluctuations.~\cite{rf:sumrule}  
 Moreover, the manifestation of the quantum fluctuations is 
pointed out from the violation of the f-sum rule.~\cite{rf:sumrule}  
 It is natural that the quantum fluctuations develop with under-doping. 
 However, we think that the more detailed discussion is necessary 
for the interpretation of the experimental results, since the contribution 
from the incoherent process violates the f-sum rule.

 Finally, we show the results for the London penetration depth. 
 The London penetration depth is related with the imaginary part of the 
conductivity 
as $1/4 \pi \lambda^{2} = \omega {\rm Im} \sigma_{{\rm tot}}(\omega) 
\mid_{\omega \to 0}$. 
 This quantity is proportional to the superfluid density and the phase 
stiffness. Therefore, the London penetration depth characterizes the 
electrodymanics of the superconductor. 
 Here, we calculate the London penetration depth when the system goes into the 
superconducting state from the pseudogap state. 
 In other words, we investigate the properties of the superconducting 
transition which is suppressed by the fluctuations. 
 The expression for the London constant $1/4 \pi \lambda^{2}$ is derived 
by subtracting the paramagnetic term $\propto \sum v^{2}(G G + F F)$ from the 
diamagnetic term $\propto \sum v^{2}(G G - F F)$. 
 Here, the in-plane and {\it c}-axis London penetration depth is expressed 
on the basis of the same approximation.

\begin{eqnarray}
  \frac{1}{4 \pi \lambda_{{\rm ab}}^{2}} = 2 \frac{e^{2}}{d} 
  T \sum_{\mbox{\boldmath$k$},{\rm i} \omega_{n}} 
  v^{2}(\mbox{\boldmath$k$})
  |{\mit{\it F}} (\mbox{\boldmath$k$},{\rm i} \omega_{n})|^{2},
\\
  \frac{1}{4 \pi \lambda_{{\rm c}}^{2}} = 8 d e^{2} 
  T \sum_{\mbox{\boldmath$k$},{\rm i} \omega_{n}} 
  t_{\perp}^{2}(\mbox{\boldmath$k$})
  |{\mit{\it F}} (\mbox{\boldmath$k$},{\rm i} \omega_{n})|^{2}.
\end{eqnarray}

 The contribution from the incoherent process does not exist because of the 
$d$-wave symmetry and the momentum independence of the tunneling matrix. 
 Actually, the incoherent process contributes in the under-doped region 
by the momentum dependence of the tunneling matrix.~\cite{rf:incoherent}

\begin{figure}[htbp]
  \begin{center}
   \epsfysize=6cm
$$\epsffile{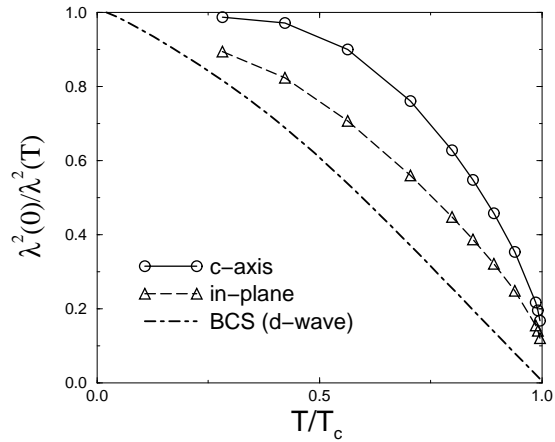}$$
%   \epsfile{file=superfluid.eps,height=6cm}
    \caption{The temperature dependence of $\lambda^{2}(0)/\lambda^{2}(T)$. 
             The open circles and the open triangles show the results along 
             the {\it c}-axis and along the {\it ab}-axis, respectively. 
             The dash-dotted line shows a result of the $d$-wave BCS theory. 
             Here, $\lambda(0)$ is derived by the extrapolation. 
             $1/\lambda^{2}(T)$ grows rapidly near $T_{{\rm c}}$ with 
             the rapid growth of the order parameter. 
             The rapid growth is more remarkable along the c-axis. 
             }
  \end{center}
\end{figure}

 We show the temperature dependence of the in-plane and {\it c}-axis 
London penetration depth in Fig. 12. 
 The inverse square of the London penetration depth $1/\lambda^{2}$ has almost 
linear temperature dependence within the $d$-wave BCS theory. 
 The linear dependence in the low temperature region is 
due to the quasi-particle excitation near the gap node. 
 Near the critical temperature $T_{{\rm c}}$, 
the temperature dependence of the order parameter 
$\Delta \propto (T_{{\rm c}}-T)^{1/2}$ causes the linear dependence in the 
BCS theory. 
 Our calculation shows the more rapid increase of $1/\lambda^{2}$ 
near $T_{{\rm c}}$ 
owing to the rapid growth of the order parameter. 
 The rapidness is more remarkable along the {\it c}-axis because the 
quasi-particle excitation near the gap node does not contribute. 
 In other words, the paramagnetic current is rapidly suppressed along the 
{\it c}-axis. 
 As a result, the temperature dependence of $1/\lambda_{{\rm c}}^{2}$ is 
rather weak than that of $1/\lambda_{{\rm ab}}^{2}$ at the low temperature. 
 The expression of $t_{\perp}(\mbox{\boldmath$k$})$ in eq.(5.1) 
causes $T^{5}$-law, 
$\lambda_{{\rm c}}^{2}(0)/\lambda_{{\rm c}}^{2}(T) = 1 - a T^{5} $ in the low 
temperature region.~\cite{rf:xiang} 
 These behaviors are qualitatively consistent with the experiment 
for the several under-doped or optimally-doped 
cuprates,\cite{rf:xiang,rf:penetrate} 
although our calculation can not expect the precise temperature dependence 
near the critical point.

\section{Summary and Discussion}

 In this paper, the pseudogap state and the superconducting state in 
High-$T_{{\rm c}}$ cuprates have been investigated. 
 In particular, we have paid attention to the close relation between the 
pseudogap state and the superconducting state. 
 We have given the results calculated for various physical quantities 
which show the pseudogap anomaly. 
 We have started from the model with the important factors in realizing 
the pseudogap.~\cite{rf:yanasePG} 
 The calculation is based on the self-consistent T-matrix 
approximation which we have used before in order to explain the pseudogap 
phenomena. 
 The effects of the superconducting fluctuations are included in the 
self-energy correction. 
 It should be noticed that the system is not the low density system. 
 Therefore, our scenario is based on the resonance scattering and 
is not based on the Nozi$\grave{{\rm e}}$res and Schmitt-Rink 
theory.~\cite{rf:Nozieres}  
 Namely, the mechanism of the phase transition should be considered as 
the superconductivity, and is not the Bose condensation 
in the sense of the NSR theory. 
 The characteristics of the pseudogap state and the superconducting state have 
been investigated on the basis of the calculated results. 
 The calculated results well explain the experimental results. 
 The smooth change observed in the electronic spectrum has been reproduced. 
 The same energy scale is not so self-evident in the resonance scattering 
scenario, compared with the NSR theory. 
 However, our results have actually confirmed the important character. 
 This fact means that the energy gain due to the formation of the gap around 
the Fermi energy is common to both of the pseudogap and the superconductivity.

 Moreover, most of the pseudogap phenomena in the various measurements 
originate in the change of the single particle properties. 
 We have calculated the magnetic and transport quantities. 
 The results for the NMR $1/T_{1}T$, $1/T_{2G}$, neutron scattering, in-plane 
and {\it c}-axis optical conductivity, and the London penetration depth have 
been shown. 
 The comprehensive understanding of their behaviors has been given 
both in the pseudogap state and in the superconducting state.

 Generally speaking, the comprehensive explanation of the phase diagram is 
important to understand the overview of the High-$T_{{\rm c}}$ 
superconductivity. 
 The pseudogap is an especially important character of the High-$T_{{\rm c}}$ 
cuprates. 
 The following two points are important in the comprehensive understanding 
of the pseudogap. 

 One is the continuity with respect to the hole doping. 
The pseudogap phenomena continuously take place from slightly over-doped to 
under-doped region. 
 The pseudogap is weak in the slightly over-doped or optimally-doped region,  
while the strong pseudogap behavior appears in the under-doped region. 
 Our scenario based on the strong coupling superconductivity is properly 
consistent with the doping dependence. 
 As the system approaches the half-filled Mott insulator, 
the pairing interaction via the anti-ferromagnetic spin fluctuations 
increases and the effective Fermi energy $\varepsilon_{{\rm F}}$ 
is renormalized. 
 Therefore, the superconducting coupling becomes strong with under-doping, 
and the pseudogap phenomena become strong. 
 Since the energy scale in our calculation is scaled by the effective Fermi 
energy $\varepsilon_{{\rm F}}$, the doping dependence of $T_{{\rm c}}$ is 
well explained.~\cite{rf:yanasePG}  
 The typical change due to the doping is observed in the magnetic field 
dependence of the pseudogap 
phenomena.~\cite{rf:zheng,rf:gorny,rf:mitrovic,rf:eschrig,rf:zheng2}  
 The scenario based on the resonance scattering well explains the doping 
dependent character of the pseudogap continuously from the slightly 
over-doped to under-doped region.~\cite{rf:yanaseMG}  
 The comprehensive understanding strongly supports the scenario. 

 The other important point is the close relation between the pseudogap state 
and the superconducting state. 
 It is natural to consider the pairing scenario for the pseudogap which 
develops with approaching the critical point and smoothly changes to the 
superconducting gap. 
 In this paper, the same aspects between the pseudogap state and the 
superconducting state are confirmed in detail. 
 The results support our scenario.

 Here, we briefly discuss some remained problems. 
 The in-plane transport in the pseudogap state is considered to be explained 
by the feedback effect on the low frequency spin 
fluctuations.~\cite{rf:yanaseTR,rf:dahmtransport}  In particular, the behavior 
of the Hall coefficient is interesting. 
 The vertex correction due to the spin fluctuations plays an important role 
for the strong enhancement of the Hall coefficient.~\cite{rf:kontani} 
 The vertex correction is also reduced by the feedback effect on the spin 
fluctuations. 
 Therefore, the Hall coefficient is expected to be reduced by the pseudogap. 
 The reduction is actually observed in the experiments.~\cite{rf:transport} 
 Moreover, the other effects on the electronic state exist. Since the vertex 
correction is important at 'hot spot' where the pseudogap is large,  
 the pseudogap itself reduces the vertex correction. 
 The detailed calculation is an important future problem.

 In this paper, we have investigated the temperature dependence of the London 
constant $\Lambda(T) \propto 1/\lambda(T)^{2}$. 
 We think that the low temperature behavior of the penetration depth is 
interesting as a typical problem of the superconductivity in the strongly 
correlated electron 
systems.~\cite{rf:randeriaphase,rf:uemura,rf:millis,rf:mesot,rf:legget} 
 In particular, the strongly anisotropic High-$T_{{\rm c}}$ cuprates are 
expected to have the characteristic temperature and doping 
dependences.  The Uemura plot is a typical one.~\cite{rf:uemura} 
 Moreover, the inconsistency of the absolute value $\Lambda(0)$ 
and the coefficient of the T-linear term of $\Lambda(T)$ has been pointed out 
on the analogy of the result for the isotropic 
system.~\cite{rf:millis,rf:mesot,rf:randeriaphase}  
 The absolute value $\Lambda(0)$ decreases with under-doping, 
while the coefficient of the T-linear term is almost independent of 
the doping.~\cite{rf:millis} 
 We consider that the inconsistency is understood by considering the 
momentum dependence of the current vertex $J_{\nu}$. 

 The current vertex $J_{\nu}$ 
includes the effects of the mass renormalization 
and the back flow.~\cite{rf:legget,rf:okabe,rf:maebashi}  
 The absolute value $\Lambda(0)$ is determined by the total quasiparticles 
near the Fermi surface, while the T-linear term is determined 
by the quasiparticles near the gap node. 
 The London constant is reduced by the mass renormalization which is 
is large near $(\pi,0)$ and is small near $(\pi/2,\pi/2)$.~\cite{rf:yanaseTR} 
 The cancellation due to Fermi liquid effect does not occur in the 
anisotropic systems.~\cite{rf:legget,rf:okabe,rf:maebashi}  
 Therefore, the London constant $\Lambda(0)$ 
is roughly scaled by the effective Fermi energy $\varepsilon_{{\rm F}}$. 
 By considering that the critical temperature $T_{{\rm c}}$ is also scaled 
by the effective Fermi energy $\varepsilon_{{\rm F}}$ 
in the strong coupling region,~\cite{rf:yanasePG} 
we can understand the Uemura plot 
$T_{{\rm c}} \propto \Lambda(0) $ ~\cite{rf:emery,rf:uemura}
observed in the under-doped region. 
 We consider that the vertex correction (that is to say, the Fermi 
liquid effects) reduces the current vertex furthermore. 
 By considering the Bethe-Salpeter equation for the current vertex, 
we can see the current vertex is reduced by the vertex correction 
exchanging the anti-ferromagnetic spin fluctuations. 
 The vertex correction is different from that for calculating 
the Hall coefficient. 
 The correction for the London constant is related to the real part of 
$\chi_{{\rm s}}(\mbox{\boldmath$q$},\omega)$, 
while that for the Hall coefficient is related to the imaginary part. 
 Therefore, the effect remains even in the superconducting state. 
 The vertex correction is also large at `hot spot' and small at `cold spot'. 
 Therefore, the Fermi liquid effect makes the system more anisotropic. 
 As a result, the strong correlation reduces the absolute value, while the 
T-linear term is not reduced so much. 
 The detailed calculation is an important future problem.

\section*{Acknowledgements}

 The authors are grateful to Dr. S. Koikegami for fruitful discussions, 
and to Dr. S. Fujimoto for some fruitful comments. 
 Numerical computation in this work was partly carried out 
at the Yukawa Institute Computer Facility. 
 The present work was partly supported by a Grant-In-Aid for Scientific 
Research from the Ministry of Education, Science, Sports and Culture, Japan. 
 One of the authors (Y.Y) has been supported by a Research Fellowships of the 
Japan Society for the Promotion of Science for Young Scientists.

\end{document}